\documentclass[a4paper,11pt]{article}
\pdfoutput=1 

\usepackage{jheppub} 

\usepackage{slashed}

\title{\boldmath Additional Information on Heavy Quark Parameters from Charged
Lepton Forward-Backward Asymmetry}

\author[a]{Sascha Turczyk}

\affiliation[a]{PRISMA Cluster of Excellence \& Mainz Institute for Theoretical
Physics, Johannes Gutenberg University, 55099 Mainz, Germany}

\emailAdd{turczyk@uni-mainz.de}

\abstract{The determination of $|V_{cb}|$ using inclusive and exclusive
(semi-)leptonic decays exhibits a long-standing tension of varying ${\cal
O}(3 \sigma)$ significance. For the inclusive determination the decay rate is
expanded in $1/m_b$ using heavy quark expansion, and from moments of physical
observables the  higher order heavy quark parameters are extracted from
experimental data in order to assess $|V_{cb}|$ from the normalisation. The drawbacks are high correlations both theoretically as
well as experimentally among these observables. We will scrutinise the inclusive
determination in order to add a new and less correlated observable. This
observable is related to the decay angle of the charged lepton and can help to
constrain the important heavy quark parameters in a new way. It may validate
the current seemingly stable extraction of $|V_{cb}|$ from inclusive decays or
hints to possible issues, and even may be sensitive to New Physics operators.}

\begin{document} 
\begin{flushright}
MITP/15-110 
\end{flushright}
\maketitle

\section{Introduction}

The cleanest way to access the matrix elements of  the
Cabibbo-Kobayashi-Maskawa (CKM) matrix are (semi-)leptonic
decays~\cite{Agashe:2014kda}. Besides precise experimental data a reliable
theoretical framework is necessary. Heavy Quark Symmetry (HQS) and Heavy Quark
Expansion (HQE) have proven to be very successful in describing decays of heavy
$B$-mesons~\cite{HQE0,HQE1,HQE2,HQE3}. Especially in the case of inclusive
semi-leptonic decays there has been great effort both from
experiment~\cite{Aubert:2009qda,Aubert:2004td,Urquijo:2006wd,Schwanda:2006nf,
Schwanda:2008kw,Acosta:2005qh,Csorna:2004kp,Abdallah:2005cx} and
theory~\cite{U1,U2,S1,S2} to push the precision of $|V_{cb}|$ down to ${\cal
O}(\lesssim 2\%)$ in the global fit~\cite{Alberti:2014yda}. The HQE is a double
expansion in $1/m_b$ and $\alpha_s$. Current state-of-the-art analysis are
theoretical calculations up to  ${\cal O}(\alpha_s^2)$
\cite{Melnikov:2008qs,Pak:2008qt}, the mixed ${\cal O}(\alpha_s / m_b )$~
\cite{Becher:2007tk,Alberti:2012dn,Alberti:2013kxa,Mannel:2014xza,
Mannel:2015jka} and ${\cal O}(1/m_b^5)$ non-perturbative corrections; the
results for
$1/m_b^3$ have been known~\cite{Blok:1993va,Gremm:1996df} for quiet some time,
while the calculations of $1/m_b^4$ ~\cite{Dassinger:2006md} and $1/m_b^5$
~\cite{Mannel:2010wj} including investigations concerning subtleties due to the
heavy final state quark ``intrinsic-charm''
\cite{Bigi:2005bh,Breidenbach:2008ua,Bigi:2009ym} have lately been performed. 
The most recent global fit~\cite{Alberti:2014yda} uses the theoretical
calculations up to $1/m_b^3$ and all known radiative corrections, however the
fit results for the extracted $|V_{cb}|$ seem to be rather stable under adding
higher order theoretical corrections as can be seen from older global
analysis~\cite{Bauer:2004ve,Buchmuller:2005zv,Gambino:2013rza}. The number of
new parameters at order $1/m_b^{4,5}$ proliferates, and hence these results
cannot be simply implemented into the fit to experimental data. Some numerical
studies about the effects and possible extraction of some of the most important
parameters are ongoing~\cite{GambinoTurczyk:InPreparation}, using partially
estimates of these higher-order matrix elements~\cite{Heinonen:2014dxa}.

The higher precision and especially accumulated data of the future Belle-II
experiment~\cite{Aushev:2010bq} will be able to make use out of this additional and less correlated
observable. It will hopefully help to disentangle the tension with respect to
the extraction of $|V_{cb}|$ utilising other methods or may contribute to solve some other 
puzzles in this decay mode~\cite{Bernlochner:2012bc}. It has been noted before~\cite{Buras:2010pz,Crivellin:2009sd}, that right-handed currents may help to ease this tension, especially in
$b\rightarrow u$ transitions. The relation between the transition to the heavy charm quark or light up quark is however
model-dependent. First studies have estimated the potential impact of
right-handed currents both in inclusive~\cite{Dassinger:2007pj,Dassinger:2008as,Feger:2010qc}
as well as exclusive $b\rightarrow c \ell \bar{\nu}_\ell$ transitions~\cite{Faller:2011nj}, still allowing for a few percent of a right-handed current admixture. A recent LHCb
analysis~\cite{Aaij:2015bfa} using a baryonic decay mode disfavours this New Physics (NP) interpretation of the tension in $b\rightarrow u$ transitions\footnote{Note that this measurement of $|V_{ub}|$ depends on the value of $|V_{cb}|$, which has been fixed to the value extracted from exclusive semi-leptonic decays. Using the inclusive value for $|V_{cb}|$, the extracted central value of $|V_{ub}|$ would be larger by about 7\%.}, while there has been a possible solution prior to this measurement~\cite{Bernlochner:2014ova}. 

We believe the statement of \cite{Crivellin:2014zpa} that right-handed currents in $b\rightarrow u(c)$ semi-leptonic transitions are already ruled out by
data is too strong. Their motivation to exclude right-handed currents in $b\rightarrow u$ transitions bases purely on the reinterpretation of the $B\rightarrow \rho \ell \bar{\nu}_\ell$ measurement. Besides issues with experimentally identifying the broad $\rho$-resonance~\cite{Faller:2013dwa,Kang:2013jaa} in accordance with its theory description especially for the normalisation,  their reinterpretation of experimental data integrated over a range of $q^2$ into a single value of $q^2$ outside of this region does neither take into account efficiency corrections of the altered $q^2$ spectrum due to NP contributions, nor uses it the theoretical non-perturbative predictions at a point in phase-space, where these are valid and the uncertainties are trustworthy. Therefore neither the central value nor the uncertainty band as a function of the right-handed admixture are computed reliably. Hence their conclusion to exclude right-handed currents purely to a
deviation from their derived uncertainty band at the one sigma level is too strong. In a first order approximation in~\cite{Bernlochner:2014ova} we have taken into account such effects to reinterpret the Neyman belt using the same experimental data, in the valid theoretical range of low $q^2$ for the form factor predictions. In this
analysis right-handed currents may not be excluded, yet. It is obvious, that a correct exclusion calls for a revisit of the measurements with taking into account efficiency and acceptance corrections for the NP altered spectrum. The statement for the $b\rightarrow c$ transition is less severe. Furthermore even in the
purely exclusive extraction there exist still a discrepancy using either light-cone sum rules or lattice QCD~\cite{Agashe:2014kda}. Hence there cannot be
a conclusive decision made with the current theoretical and experimental situation, and therefore we think the line of argumentation in  \cite{Crivellin:2014zpa} is too restrictive.

We neglect lepton masses in the following discussion. The paper is
organised as follows. In section~\ref{sec:decayrate}, we will derive
the differential spectrum and the forward-backward asymmetry and discuss
subtleties due to introducing a cut of the minimum energy for the charged
lepton. Section~\ref{sec:afbanalytic} 
provides the expressions up to order ${\cal O}(1/m_b^3)$ to demonstrate the use
and additional information of this observable in comparison to moments of the
hadronic invariant mass and charged lepton energy. In
Section~\ref{sec:afbnumerical} we will
discuss the impact of higher-orders numerically, where full results are
available analytically, and conclude in Section~\ref{sec:discussion}.

\section{Decay Rate}\label{sec:decayrate}

\subsection{Decay Kinematics}
For the following discussion to be useful in analysis, we assume that the full event
kinematics may be reconstructed experimentally. That can be
achieved at (Super-)$B$-factories~\cite{Aushev:2010bq} with hadronic tag analysis to reconstruct the kinematics including the invisible neutrino momentum. This decay kinematics is given by 
\begin{equation}
  p_B^\mu = p_\ell^\mu + p_{\bar\nu_\ell}^\mu + p_{x_C}^\mu \,{:}{=}\, q^\mu + p_{x_C}^\mu\,.
\end{equation}
Here $q^\mu$ is the
momentum transfer to the lepton system, and $p_B^\mu = m_B v^\mu$ with $v^\mu$ being the four velocity of the
$B$-meson, and $(v^\mu)=(1,0,0,0)$ in the $B$-meson rest-frame. We calculate the fully
differential rate in the three kinematical variables
\begin{subequations}\label{Def:Kinematics}
\begin{align}
  q^2 &= 2 p_\ell \cdot p_{\bar \nu_\ell} \label{Def:q2}\\
  v{\cdot}q &= v {\cdot}  p_\ell + v{ \cdot} p_{\bar \nu_\ell} \label{Def:vq}\\
  z \,{:=}\, \cos \theta &= \frac{v {\cdot} p_{\bar \nu_\ell} -v{ \cdot}  p_\ell   }{\sqrt{v
{\cdot} q^2-q^2}} \label{Def:theta}\,.
\end{align}
\end{subequations}
The angle $z=\cos \theta$ is defined the same way as for the forward-backward
asymmetry~\cite{Ali:1991is} $A_{FB}$ in the flavor changing neutral current
decay $b \rightarrow s \ell^+
\ell^-$:
it is given by the angle of the charged lepton with the flight direction of the
$B$-meson, in the rest-frame of the lepton-anti-neutrino system ($\vec q = 0$). As given in
Eq.~\eqref{Def:theta} it can be related to the energies of both leptons and the
momentum transfer to the lepton system in the $B$-rest-frame. In this form it can be seen that $z$ is a Lorentz invariant observable. All other possible contractions of appearing four momentum vectors depend
linearly on the choice of \eqref{Def:Kinematics}.

\subsection{Differential Decay Rate}
The differential rate can be decomposed into the leptonic and hadronic tensor
\begin{equation}
  \text{d}\Gamma = 16 \pi G_F^2 |V_{cb}|^2 W_{\mu \nu} L^{\mu \nu}
\text{d}\phi\,,\label{Eq:dGamma}
\end{equation}
where we have defined the leptonic and hadronic tensor as
\begin{align}
  L^{\mu \nu} &= \sum_{\underset{\text{spins}}{\text{\tiny lepton}}} \langle 0 |
J^{\nu,\dagger}_\ell | \ell \bar{\nu_\ell}\rangle \langle \ell \bar{\nu_\ell} |
J^{\mu}_\ell | 0 \rangle \\
  W_{\mu \nu} &= \frac{1}{2m_B} \sum_{X_c} \langle \bar B | J^\dagger_{q,\nu} |
X_c \rangle \langle X_c | J_{q,\mu} | \bar B\rangle (2 \pi)^3 \delta^{(4)} (p_B
- (p_\ell + p_{\bar \nu_\ell} + p_{X_c}))\,,
\end{align}
and used the abbreviation for the Standard Model (SM) quark and hadronic current
\begin{align}
  J^{\mu}_\ell &= \bar\ell  \gamma^\mu \frac{1-\gamma^5}{2}{\nu_\ell} \\
  J_{q,\mu} &= \bar c \gamma_\mu \frac{1-\gamma^5}{2} b\,.
\end{align}
In case of a previously mentioned right-handed admixture, we would have
\begin{align}
  J_{q,\mu}^\text{NP} &= \bar c \gamma_\mu \frac{1-\gamma^5}{2} b  + \epsilon_R\,\, \bar c \gamma_\mu \frac{1+\gamma^5}{2} b  \,,
\end{align}
and need to redefine $V_{cb} \rightarrow V_{cb}^L$. As far as the decay kinematics is concerned, we may use the three invariants $v{\cdot}
q$, $q^2$ as well as $z = \cos\theta$. The latter one has
not been considered in tree-level decays, yet. It has been used in flavor
changing neutral currents though~\cite{Ali:1991is}, in which New Physics is suspected to show off first as it may enter at the same order as the Standard Model contribution and both leptons are charged and thus experimentally visible. As we will later see,
we treat the hadronic part in heavy quark effective theory (HQET). Then the $b$-quark momentum is given by $p_b^\mu = m_b v^\mu + k^\mu$, where the soft vector $k^\mu$ describes the off-shellness of the heavy quark. Basically we 
expand the hadronic tensor in powers of $k^\mu$, using a background field method with $k^\mu\rightarrow i D^\mu$, in a systematic way in order to preserve the correct ordering \cite{Dassinger:2006md}. Thus we treat the decay phase-space at partonic level kinematics, i.e.
\begin{equation}
  p_b^\mu = m_b v^\mu + k^\mu = q^\mu + p_c^\mu \label{Def:PartMom}\,.
\end{equation}
The off-shellness of the bottom quark will be mimicked by derivatives of the
on-shell delta distribution condition of the hadronic tensor, which occurs at
higher orders in the $1/m_b$ expansion.
This however, has no impact on the leptonic side, as we can factorise the decay
rate according to Eq.~\eqref{Eq:dGamma} and we have
\begin{subequations}\label{Def:Decomp}
\begin{align}
     \text{d}\Gamma &= 16 \pi G_F^2 |V_{cb}|^2  \tilde  W_{\mu \nu} \tilde
L^{\mu \nu} \\
      \tilde  W_{\mu \nu} &= \int \frac{\text{d}s}{2\pi} \frac{\text{d}^4
q}{(2\pi)^4} 2\pi \delta(s-q^2) W_{\mu \nu}\nonumber \\
	      &= \frac{1}{8 \pi^3} \sqrt{v{\cdot}q^2-q^2} W_{\mu \nu}
\,\,\text{d}q^2 \text{d}v{\cdot}q \label{Def:HadTFull} \\
	      \tilde L^{\mu \nu}&= \int \frac{d^3 p_\ell}{(2\pi)^3 2
E_\ell}\frac{d^3 p_{\bar \nu_\ell}}	{(2\pi)^3 2 E_{\bar \nu_\ell}} L^{\mu \nu}
(2\pi)^4 \delta^{(4)} (q-p_\ell - p_{\bar \nu_\ell})\,.\label{Def:LepTFull}
\end{align}
\end{subequations}
The hadronic tensor can be decomposed into structure functions depending each only on
$v{\cdot}q$ and $q^2$
\begin{equation}
 	W_{\mu \nu} = -g_{\mu \nu} W_1 + v_\mu v_\nu W_2 - i \epsilon_{\mu \nu
\alpha \beta} v^\alpha q^\beta W_3 + q_\mu q_\nu W_4 + (v_\mu q_\nu + v_\nu
q_\mu) W_5\,. \label{Def:HadT}
\end{equation}
The leptonic tensor is simply given by
\begin{equation}
 	L^{\mu \nu} = 2\left( p_\ell^\mu p_{\bar\nu_\ell}^\nu +p_\ell^\nu
p_{\bar\nu_\ell}^\mu-g^{\mu \nu} p_\ell \cdot p_{\bar\nu_\ell} -i \epsilon^{\mu
\nu \eta \lambda } p_{\ell \eta} p_{\bar\nu_\ell \lambda}
\right)\,.\label{Def:LepT}
\end{equation}
The contraction of the leptonic tensor and hadronic tensor in turn
is then given by
\begin{equation}
  L^{\mu \nu} W_{\mu \nu} =2 q^2 \,W_1 + \left(1-z^2 \right)\left(v{\cdot}q^2 -q^2\right)\,W_2  +2 q^2 z \sqrt{v{\cdot}q^2-q^2}\,W_3 \label{Eq:FullContract}\,.
\end{equation}
As can be seen from this equation, the contribution from $W_3$ is sensitive to asymmetric integrations over $z$, which for example is true in observables as a forward-backward asymmetry, while the other two terms with $W_{1,2}$ drop out. In contrast for the regular integration over the whole kinematically allowed region of $z$, as done for all ``conventional'' observables, the contribution from $W_3$ drops out and we are purely sensitive to $W_{1,2}$. Hence we are interested in constructing the observable such, that we gain additional information on $W_3$, which is otherwise lost.

The decomposition we have elaborated on in Eq.~\eqref{Def:Decomp} enables us to calculate the phase-space for the triple differential decay rate in the following subsection.

\subsection{Phase-Space Integration for Forward-Backward Asymmetry}
By construction the dependence of $v{\cdot}q$ and $q^2$ is contained in the hadronic tensor. We
need to perform the phase-space integration over the leptonic degrees of freedom including the leptonic tensor with implicitly retaining the dependence
on the angle $z$. Strictly speaking, the phase-space integration is only valid for the full contraction of the hadronic tensor with the 
leptonic tensor given by Eq.~\eqref{Eq:FullContract}, which we keep in mind in
the following\footnote{Alternatively we may decompose $I_{\mu \nu} (v{\cdot} q, q^2, z) = \int \text{d}\phi L_{\mu \nu}$ into leptonic structure functions. However then we were not be able to identify the $z$ dependence, which multiplies the hadronic structure functions in Eq~\eqref{Eq:FullContract}, because it is only contained in $I_{\mu \nu}$ due to its relation with the leptonic phase-space.}. This contraction does depend only on the three kinematic variables in Eq.~\eqref{Def:Kinematics}, and due to the hadronic part in
Eq.~\eqref{Def:HadTFull} we are already differential in $v{\cdot}q$ and $q^2$.
We calculate the phase-space for massless leptons, and we introduce the
dependence on the angular variable explicitly
\begin{align}
  \int \text{d}\phi &= \int \frac{\text{d}^3 p_\ell}{(2\pi)^3 2 E_\ell} \frac{\text{d}^4 p_\nu}{(2\pi)^4} (2\pi) \delta(p_\nu^2) (2\pi)^4 \delta (q-p_\nu - p_\ell) \theta(p_\nu^0) \nonumber \\
    &\phantom{= \int } \,\times \text{d}z\, \delta \bigg (\!\!z - \frac{p_\nu^0 - p_\ell^0}{\sqrt{v{\cdot}q^2-q^2}}\!\bigg ) \theta (E_\ell - E_\text{cut}) \nonumber \\
    &= \int \frac{\text{d}\Omega_\ell}{(2\pi)^2} \frac{E_\ell^2\, \text{d}E_\ell}{2 E_\ell} \text{d} z \,\delta\bigg( (q-p_\ell)^2 \bigg) \,\delta \bigg (\!\!z -\frac{v{\cdot}q - 2 E_\ell}{\sqrt{v{\cdot}q^2-q^2}}\! \bigg )  \theta (E_\ell - E_\text{cut}) \theta(v{\cdot}q - E_\ell) \nonumber \\
    &= \int \frac{\text{d}\cos \theta_\ell}{2 \pi} \frac{\text{d} E_\ell}{4 \sqrt{ v{\cdot}q^2 - q^2}} \delta \bigg( \!\!\cos \theta_\ell - \frac{2 E_\ell v{\cdot}q - q^2}{2 E_\ell \sqrt{ v{\cdot}q^2 - q^2}} \!\bigg) \theta(1+ \cos \theta_\ell) \theta(1-\cos\theta_\ell)   \nonumber \\
      &\phantom{= \int } \,\times \text{d}z\,\frac{\sqrt{v{\cdot}q^2-q^2}}{2}\delta \bigg (\!\!E_\ell -\frac12 \big(v{\cdot}q - z\sqrt{v{\cdot}q^2-q^2}  \big)\! \bigg ) \theta (E_\ell - E_\text{cut}) \theta(v{\cdot}q - E_\ell) \nonumber \\
      &= \frac{\text{d}z}{16\pi} \theta(q^2) \theta (v{\cdot}q^2  - q^2)\theta (v{\cdot}q - z \sqrt{ v{\cdot}q^2 - q^2} - 2 E_\text{cut}) \,. \label{Eq:PhaseSpace}
\end{align}
The angle $\cos \theta_\ell$ shall not be confused with the observable $z=\cos \theta$. In deriving this result, we have used
\begin{align}
   (q-p_\ell)^2  &= q^2 -2 q {\cdot} p_\ell \nonumber \\
	    &= q^2 -2 (v {\cdot} q E_\ell - |\vec q| |\vec p_\ell| \cos \theta_\ell) \nonumber \\
	    &= q^2 -2 v{\cdot}q E_\ell + 2 E_\ell \sqrt{ v{\cdot}q^2 - q^2} \cos \theta_\ell \nonumber \\
	    \Rightarrow \qquad  \delta\bigg( (q-p_\ell)^2 \bigg) &= \frac{1}{2 E_\ell \sqrt{ v{\cdot}q^2 - q^2}} \delta \bigg( \!\!\cos \theta_\ell - \frac{2 E_\ell v{\cdot}q - q^2}{2 E_\ell \sqrt{ v{\cdot}q^2 - q^2}} \!\bigg)\,.
\end{align}
In the last step we have evaluated the integrals using the delta distributions. For applying this to the lepton angle, we needed to introduce further theta distributions to limit the integration region of $\cos \theta_\ell$ to the physical ones. Then we have simplified the kinematical constraints of the theta distributions, and we will later see that 
these distributions are necessary for the derivation of integrated observables. Trivial conditions may be neglected. 

In summary the triple differential decay rate is written as
\begin{align}
  \frac{\text{d}^3 \Gamma}{\text{d}v{\cdot}q\, \text{d} q^2\, \text{d}z} =  &\frac{G_F^2 | V_{cb}|^2}{192 \pi^3 m_b^5} \,24 m_b^5 \sqrt{ v{\cdot}q^2 - q^2}  \nonumber \\
      &\times\left[2\, q^2 \,W_1 + \left(1-z^2\right)\left(v{\cdot}q^2 -q^2\right)W_2  +2\, z \,q^2\,  \sqrt{v{\cdot}q^2-q^2}\,W_3 \right] \nonumber \\
       &\times\theta(q^2) \theta (v{\cdot}q^2  - q^2)\theta (v{\cdot}q - z \sqrt{ v{\cdot}q^2 - q^2} - 2 E_\text{cut})\,. \label{Eq:TripleRate}
\end{align}

\subsection{Hadronic Tensor in Heavy Quark Expansion}
We proceed along the lines of \cite{Mannel:2010wj} to compute the hadronic tensor in the HQE, which we shall briefly summarise here. We start with a non-local forward matrix element of the form
\begin{equation}
 	T_{\mu \nu} = -\frac{i}{2M_B} \int \text{d}^4x e^{-i qx} \langle \bar B| \text{T} \big[J^\dagger_{q, \nu}(x),J_{q,\mu}(0)  \big] |\bar B\rangle\,.
\end{equation}
\begin{figure}[htp]
    \centering\includegraphics[scale=1]{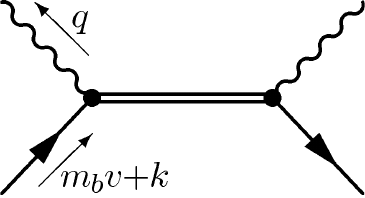}\caption{Background field propagator.}\label{fig:feynman_tree}
\end{figure}
This can be visualised by the Feynman diagram in Fig.~\ref{fig:feynman_tree}.
The double line denotes the charm quark that propagates in the soft background
fields of the meson. We relate this time-ordered product to the hadronic tensor
by the optical theorem
\begin{equation}
 	-\frac{1}{\pi}\,\text{Im}\,T_{\mu \nu} = \frac{1}{2M_B}  \sum_{X_c}\langle \bar B| J^\dagger_{q,\nu} |X_c \rangle\langle X_c|J_{q,\mu} |\bar B \rangle (2\pi)^3\delta^4 (p_B  - q - p_{X_c}) = W_{\mu \nu}\,.
\end{equation}
The soft momentum $k$ of the momentum transfer $p_b-q$ from
Eq.~\eqref{Def:PartMom} is replaced by the covariant derivative in the charm
quark propagator, containing the soft background field gluons. This propagator
then becomes the background field (BGF) propagator
\begin{equation}
    S_\text{BGF} = \frac{1}{m_b \slashed{v} + i\slashed{D} - \slashed{q}  -m_c + i \epsilon}\,. \label{eq:sbgf}
\end{equation}
The BGF propagator describes the charm quark propagating in the forward matrix
element of the $B$-meson with all the soft fields, for instance binding gluons,
and therefore accounts for the difference between the partonic quark picture and
the non-perturbative nature of the meson. We write this non-local propagator as
a geometric series, to yield an expansion in $k^\mu / m_b$ with $Q^\mu = m_b v^\mu - q^\mu$
\begin{equation}
    S_\text{BGF} = \left[ \sum_{n=0}^\infty (-1)^n\left[\left(\slashed{Q}-m_c + i \epsilon\right)^{-1}(i\slashed{D})\right]^n \right]\left(\slashed{Q}-m_c\right)^{-1}
\end{equation}
and the operator product expansion (OPE) can be cut off at some mass dimension
$m$. In our case we compute up to $m=5$, which yields the expansion up to
$1/m_b^5$. Notice that the application of the optical theorem
\begin{equation}
 	-\frac{1}{\pi}\text{Im}\,T_{\mu \nu} = W_{\mu\nu}\,,\label{eq:OT}
\end{equation}
can be evaluated explicitly by the means of
\begin{equation}
	-\frac{1}{\pi} \text{Im} \left(\frac{1}{\Delta_0}\right)^{n+1} = \frac{(-1)^n}{n!} \delta^{(n)}\left(Q^2 - m_c^2\right)\,, \label{Eq:OT}
\end{equation}
and we have defined $\Delta_0 = Q^2 - m_c^2 + i \epsilon$. Thus we find
derivatives of the on-shell condition for the higher-order terms, reassembling
the non-locality of the unexpanded BGF propagator into local terms. In this
procedure the full QCD field in the OPE is retained and we are left with only
local operators. No additional non-local pieces from expanding the state as well
as the field will occur, however the relation to other heavy hadrons containing
a heavy quark is now only true up to corrections of order $1/m_Q$ and
$\alpha_s$. To coincide with the usually defined parameters in dimension 5,
which is equal to expanding up to $1/m_b^2$, we define the operators to
be\begin{subequations}\label{eq:dim5}
\begin{align}
    2M_B\, \mu_\pi^2 &= - \langle \bar B| \bar b_v \, i D_\rho i D_\sigma  \,b_v |\bar B\rangle \,\,  \Pi^{\rho \sigma}   \label{eq:dim5Spinless}\\
    2M_B\, \mu_G^2 &= \frac12 \langle \bar B|  \bar b_v \,\big[ i D_\rho , iD_\sigma\big]  \big (-i \sigma_{\alpha \beta}\big)\,b_v  |\bar B \rangle  \,\,  \Pi^{\alpha \rho} \Pi^{\beta \sigma} \label{eq:dim5Spin} \,.
\end{align}
\end{subequations}
Here $\Pi_{\mu\nu}\equiv v_\mu v_\nu - g_{\mu\nu}$ is the projector onto the spatial components. We can identify $\mu_\pi^2$ with the kinetic energy term and $\mu_G^2$ as the chromo-magnetic moment. In dimension 6, corresponding to $1/m_b^3$ we define the Darwin term $\rho_D^3$ and the spin-orbit term $\rho_{LS}^3$ as
\begin{subequations}\label{eq:dim6}
\begin{align}
    2M_B\, \rho_D^3 &= \frac12 \langle \bar B\  \bar b_v \,\Big[ i D_\rho  , \big[i D_\sigma ,  i D_\lambda \big]\Big]  \,b_v  |\bar B \rangle  \,\,  \Pi^{\rho \lambda} v^\sigma  \label{eq:dim6Spinless}\\
    2M_B\, \rho_{LS}^3 &= \frac12 \langle \bar B\  \bar b_v \, \Big \lbrace i D_\rho, \big[i D_\sigma ,  i D_\lambda \big]\Big\rbrace  \big (-i \sigma_{\mu \nu}\big)\,b_v  |\bar B\rangle  \,\,  \Pi^{\alpha \rho} \Pi^{\beta \lambda} v^\sigma \,. \label{eq:dim6Spin}
\end{align}
\end{subequations}
There appear 9 additional parameters $m_1, \ldots, m_9$ in dimension $7$ corresponding to $1/m_b^4$, and 18 in dimension $8$, which we label $r_1, \ldots r_{18}$. Their definition may be found in~\cite{Mannel:2010wj}. 

In summary, the hadronic structure tensor is written as
\begin{align}
    W_{\mu \nu} &= 
	-\frac{1}{\pi} \text{Im}\langle B(p) | \bar{b}_v \Gamma^\dagger_\nu i S_\text{BGF}  \Gamma_\mu b_v | B(p) \rangle \nonumber \\
	&=  \sum_i  \text{Tr}  \left\{   \Gamma^\dagger_\nu (\slashed{Q}+m_c) \Gamma_\mu \, \hat{\Gamma}^{(i)}   \right\}   A^{(i,0)} \delta\!\left(Q^2 - m_c^2\right) \nonumber \\ 
 	+&  \sum_i  \text{Tr}  \left\{ \Gamma^\dagger_\nu  (\slashed{Q}+m_c) \gamma^{\mu_1}  (\slashed{Q}+m_c) \Gamma_\mu   \, \hat{\Gamma}^{(i)}  \right\}  A^{(i,1)}_{\mu_1}  \delta^{(1)}\!\left(Q^2 - m_c^2\right) \nonumber  \\ 
 	+&  \sum_i  \text{Tr} \left\{   \Gamma^\dagger_\nu  (\slashed{Q}+m_c) \gamma^{\mu_1}  (\slashed{Q}+m_c) \gamma^{\mu_2}  (\slashed{Q}+m_c)\Gamma_\mu   \hat{\Gamma}^{(i)}    \right\}    A^{(i,2)}_{\mu_1 \mu_2}\frac{\delta^{(2)}\!\left(Q^2 - m_c^2\right)}{2} \nonumber\\ 
	+& \cdots    \nonumber \\
	+&\sum_i  \text{Tr}  \left\{   \Gamma^\dagger_\nu  (\slashed{Q}+m_c) \gamma^{\mu_1}  (\slashed{Q}+m_c) \cdot\ldots\cdot (\slashed{Q}+m_c)  \gamma^{\mu_m}  (\slashed{Q}+m_c)\Gamma_\mu  \, \hat{\Gamma}^{(i)}    \right\}  A^{(i,m)}_{\mu_1 \mu_2\ldots\mu_m}\nonumber \\
	&\hspace{2cm}\times\frac{\delta^{(m)}\!\left(Q^2 - m_c^2\right)}{m!}  \,.\label{Eq:wmunu}
\end{align}
The coefficients  $A^{(i,m)}_{\mu_1 \mu_2\ldots\mu_m}$ are known analytically up to order $1/m_b^5$ ($m=5$)~\cite{Mannel:2010wj}.
In the next subsection, we will see the impact of the kinematic limits from the theta distributions in \eqref{Eq:TripleRate} for higher order terms. Therefore we can write the triple differential rate as
\begin{equation}
  \frac{\text{d}^3 \Gamma}{\text{d} v{\cdot} Q\,\text{d} Q^2\, \text{d}z} = \sum_{n=0}^5 \frac{\text{d}^3 \Gamma^{(n)}}{\text{d} v{\cdot} Q\,\text{d} Q^2\, \text{d}z} \, \delta^{(n)}\!\left(Q^2 - m_c^2\right)\,.
\end{equation}

\subsection{Integration of the Differential Rate}
For the evaluation of the on-shell condition, it is advantageous to perform a variable transformation as we have already indicated above to simplify the BGF propagator
\begin{subequations}
 \begin{align}
    Q^\mu &= m_b v^\mu - q^\mu \\
    v{\cdot}q &= m_b -v{\cdot} Q\\
    q^2 &= m_b^2 - 2m_b v{\cdot} Q + Q^2\\
    \Rightarrow \qquad v{\cdot}q^2-q^2 &= v{\cdot}Q^2-Q^2 \\
    \text{d}v{\cdot}q\,\text{d}q^2 &= \text{d}v{\cdot}Q\,\text{d}Q^2\,.
 \end{align}
\end{subequations}
Effectively $Q$ is the momentum of the final state $X_c$ system, while $q$ is the kinematics from the lepton system. Then the delta distribution in the hadronic structure functions simplifies to
\begin{equation}
  \delta^{(n)} ((m_b v-q)^2- m_c^2) \rightarrow \delta^{(n)} (Q^2-m_c^2)\,.
\end{equation}
As a consequence the delta distribution does depend only on a single variable. Hence it is easiest to perform first the integration on $v{\cdot} Q$. From \eqref{Eq:TripleRate} we find the two conditions
\begin{equation}
  \theta(q^2) \theta (v{\cdot}q^2  - q^2) = \theta(m_b^2 - 2m_b v{\cdot} Q + q^2) 
\theta (v{\cdot}Q^2-Q^2) \,,
\end{equation}
where we have neglected the effect on the lepton energy cut, which we shall
investigate below. From this it is straight-forward to compute the double
differential rate
\begin{align}
    \frac{\text{d}^2 \Gamma^{(n)}}{\text{d} Q^2\, \text{d}z} &=  \int_{\sqrt{Q^2}}^{\frac{m_b^2+Q^2}{2 m_b}} \text{d}v{\cdot} Q \, \frac{\text{d}^3 \Gamma^{(n)}}{\text{d} v{\cdot} Q\,\text{d} Q^2\, \text{d}z}  \label{Eq:DoubleRate}\,.
\end{align}
The angular spectrum is now obtained by partially integrating Eq.~\eqref{Eq:DoubleRate} in $Q^2$ to evaluate the delta distribution
\begin{equation}
  \frac{\text{d} \Gamma}{\text{d}z} = \sum_{n=0}^5 (-1)^n \frac{\text{d}^n}{\text{d}(Q^2)^n}\frac{\text{d}^2 \Gamma^{(n)}}{\text{d} Q^2\, \text{d}z} \bigg | _{Q^2=m_c^2}\,.
\end{equation}
Now as far as this spectrum is concerned, we have several choices, which we will investigate in turn.
\begin{enumerate}
      \item First we integrate to the total rate as a cross-check with the known result \cite{Mannel:2010wj} to verify our procedure
      \begin{equation}
	\Gamma = \int_{-1}^1 \text{d} z \frac{\text{d} \Gamma}{\text{d}z} \,. \label{Eq:TRIntegral}
      \end{equation}
      \item We can analyse the differential spectrum $\frac{\text{d} \Gamma}{\text{d}z}$ itself.
      \item We can construct the forward backward asymmetry
      \begin{equation}
	A_{FB} =  \frac{ \int_{-1}^0 \text{d} z \frac{\text{d} \Gamma}{\text{d}z} - \int_{0}^1 \text{d} z \frac{\text{d} \Gamma}{\text{d}z}}{\int_{-1}^1 \text{d} z \frac{\text{d} \Gamma}{\text{d}z}}\,.\label{Eq:AFBIntegral}
      \end{equation}
      \item It is possible to construct moments of the angular distribution
	\begin{equation}
	    \langle z^n \rangle_\pm = \frac{ \int_{-1}^0 \text{d} z \,z^n \frac{\text{d} \Gamma}{\text{d}z} \pm \int_{0}^1 \text{d} z  \,z^n \frac{\text{d} \Gamma}{\text{d}z}}{\int_{-1}^1 \text{d} z  \frac{\text{d} \Gamma}{\text{d}z}}\,.
	\end{equation}
\end{enumerate}
We will analyse the differential spectrum itself and the $A_{FB}$ in sections~\ref{sec:afbanalytic} and \ref{sec:afbnumerical}. Note that we do not gain much more information from the moments $\langle z^n \rangle_\pm$, as can be inferred from 
Eq.~\eqref{Eq:TripleRate}, because $z$ is a polynomial prefactor of the
structure functions and without the electron energy cut there is no additional constraint. For illustration, we will compare even and odd moments with
$m \in \mathbb N$
\begin{subequations}\label{Eq:Zmoments}
\begin{align}
     \langle z^{2 m} \rangle_+ &\propto \frac{2}{2 m+1} \left(W_1 - \frac{2}{2m+3} W_2   \right) \\ 
     \langle z^{2 m} \rangle_- &\propto  -\frac{2}{2 m+2} W_3\\ 
     \langle z^{2 m+1} \rangle_+ &\propto \frac{2 }{2 m+3} W_3 \\ 
     \langle z^{2 m+1} \rangle_- &\propto -   \frac{2}{2 m+2} \left(W_1 - \frac{2}{2m+3} \frac{2m+3}{2m+4} W_2   \right)\,.
\end{align}
\end{subequations}
It is obvious, that we cannot gain more information on $W_3$ from moments in $z$. We would generate with higher even moments in $z$ for the total rate a different linear combination for $W_1$ and $W_2$, which is a bit different but very similar for odd moments in $z$ of the forward backward asymmetry.
As experimental uncertainties are growing for the measurement of higher moments
and the theoretical correlation is large with the linear combinations being
similar for various moments, it is probably not worthwhile to study moments in $z$ in
detail. Furthermore the moments already taken into account are linear combinations of $W_{1,2}$.

\subsection{Effect of Phase-Space Cuts}
As we can see in Eq. \eqref{Eq:TripleRate}, the introduction of a charged lepton energy cut introduces a non-trivial dependence into the phase-space integration. The additional conditions
\begin{subequations}\label{Eq:CutConstraint}
\begin{align}
  0 &\leq m_b-v{\cdot}Q - z \sqrt{ v{\cdot}Q^2 - Q^2} - 2 E_\text{cut} 	\label{Eq:CutConstraintA} \\
  0 &\leq E_\text{cut} \leq \frac{m_b^2-mc^2}{2} = E_\ell^\text{max}
\end{align}  
\end{subequations}
and the already previously appearing limits
\begin{subequations}\label{Eq:OldConstraint}
\begin{align}
      Q^2 &\leq v{\cdot} Q \\
      0 &\leq m_b^2 - 2m_b v{\cdot} Q + Q^2 \\
      -1 &\leq z \leq 1
\end{align}
\end{subequations}
will restrict the allowed integration region into several parts. We first need
to split the regions for two  conditions, where in region I the constraint from
\eqref{Eq:CutConstraint} is always fulfilled, while in region II,
we have to cut into the phase-space of $v{\cdot} Q$ and $z$
\begin{subequations}\label{Eq:CutRegions}
\begin{align}
      \frac12 \left(m_b - v{\cdot} Q  - \sqrt{v{\cdot} Q^2 -
Q^2 } \right) &\geq E_\text{cut} \geq 0 &\quad &\text{Region I}
\label{Eq:CutRegionsI} \\
     \frac12 \left(m_b - v{\cdot} Q - \sqrt{v{\cdot} Q^2 - Q^2 } \right)&\leq E_\text{cut} \leq E_\ell^\text{max} &\quad &\text{Region II} \label{Eq:CutRegionsII}\,.
\end{align}
\end{subequations}
For region I we find from Eq.~\eqref{Eq:CutRegionsI} a different upper limit for
the $v{\cdot} Q$ integration than without a cut
\begin{equation}
  \sqrt{v{\cdot} Q} \leq v{\cdot} Q \leq \frac{4 E_\text{cut} m_b-m_b^2-4 E_\text{cut}^2-Q^2}{2 \left(2 E_\text{cut}-m_b\right)}\,.
\end{equation}
The constraint for the $z$ integration in region I is obviously the same as in
the case without the minimum electron energy cut, as the additional condition
depending on $z$ is always fulfilled. 

Now for region II we find from Eq.~\eqref{Eq:CutConstraintA} a minimum value for
\begin{equation}
   v{\cdot} Q \geq \frac{4 E_\text{cut} m_b-m_b^2-4 E_\text{cut}^2-Q^2}{2 \left(2 E_\text{cut}-m_b\right)}
\end{equation}
by examining the extreme values for the angle $z=\pm 1$ , while for decreasing $|z| < 1$ the condition is relaxed. We therefore find a separation in $z$, up to which we are allowed to integrate over the full phase-space $v \cdot Q \leq \frac{m_b^2+Q^2}{2 m_b}$
\begin{equation}
   -1 \leq z \leq \frac{-4 E_\text{cut} m_b+m_b^2-Q}{m_b^2-Q}\,.
\end{equation}
For the remaining integration over the the angle 
\begin{equation}
  \frac{-4 E_\text{cut} m_b+m_b^2-Q}{m_b^2-Q} \leq z \leq 1
\end{equation}
we find a maximal allowed value for
\begin{equation}
    v{\cdot} Q \leq \frac{mb -2 E_\text{cut} - |z|\sqrt{\left(m_b-2 E_\text{cut}\right){}^2+Q^2 \left(z^2-1\right)}}{1-z^2}\,.
\end{equation}
Note, that region I reduces to the usual integration in the limit $E_\text{cut}
\rightarrow 0$, while region II moves out of the allowed region and hence does
not contribute and we recover the previous case.
So in summary, we find three integration regions, where in part there are non-trivial dependencies among the integration variables. The constraints are given by
\begin{subequations}\label{Eq:Regions}
\begin{align}
    \text{Region I.A: } \qquad\qquad \qquad &\qquad \nonumber \\
    ~ \sqrt{Q^2} &\leq v{\cdot} Q \leq \frac{4 E_\text{cut} m_b-m_b^2-4 E_\text{cut}^2-Q^2}{2 \left(2 E_\text{cut}-m_b\right)} \nonumber \\
    ~  -1 &\leq z \leq 1\,.  \\
    \nonumber \\
    \text{Region II.B: } \qquad\qquad \qquad &\nonumber \\
    \frac{4 E_\text{cut} m_b-m_b^2-4 E_\text{cut}^2-Q^2}{2 \left(2 E_\text{cut}-m_b\right)} &\leq v{\cdot} Q  \leq \frac{m_b^2+Q^2}{2 m_b} \nonumber \\
    -1 &\leq z \leq \frac{-4 E_\text{cut} m_b+m_b^2-m_c^2}{m_b^2-m_c^2} \,{:=}\,  z_\text{cut}\,.\\
        \nonumber \\
    \text{Region II.C: } \qquad\qquad \qquad &\nonumber \\
    \frac{4 E_\text{cut} m_b-m_b^2-4 E_\text{cut}^2-Q^2}{2 \left(2 E_\text{cut}-m_b\right)} &\leq v{\cdot} Q  \leq \frac{mb -2 E_\text{cut} - |z|\sqrt{\left(m_b-2 E_\text{cut}\right){}^2+Q^2 \left(z^2-1\right)}}{1-z^2}\nonumber \\
    z_\text{cut} \,{:=}\,   \frac{-4 E_\text{cut} m_b+m_b^2-m_c^2}{m_b^2-m_c^2} &\leq z \leq 1  \,.
\end{align}
\end{subequations}
In the following we will restrict the cut separation in $z$
\begin{equation}
  z_\text{cut} = \frac{-4 E_\text{cut} m_b+m_b^2-m_c^2}{m_b^2-m_c^2}
\end{equation}
such, that this quantity is positive for reasons that will become obvious. Consequently we have $|z| = z$ in region II.C, which will be used below. Then
\begin{equation}
  0 \leq E_\text{cut} \leq \frac{m_b^2-m_c^2}{4 m_b} \approx 1.08 \text{ GeV}\,.
\end{equation}
For the numerical estimate we have used the latest fit results in~\cite{Alberti:2014yda}. The analysis with an even larger charged lepton energy cut would in principle be the same, however some of the contributions for the forward-backward asymmetry would shift between the positive and negative term. As a realistic cut from current analysis
is $ E_\text{cut} \lesssim 1 \text{ GeV}$ or maybe below for future
analysis\footnote{For the precision of theoretical predictions a lower cut would
be preferred, as a too large cut has an impact on the validity of the HQE.},
this is a good starting point. However the constraint should be kept in mind and
fits, if not both masses are taken from other sources as already done for the
charm quark mass~\cite{Alberti:2014yda}, should be verified afterwards to
fulfil this condition in order to check if the predictions for $A_{FB}$
actually match.

Therefore the double differential rates in the three regions are obtained by the
integrals
\begin{subequations}
\begin{align}
    \frac{\text{d}^2 \Gamma_A^{(n)}}{\text{d} Q^2\, \text{d}z} &=   \int_{\sqrt{Q^2}}^{\frac{4 E_\text{cut} m_b-m_b^2-4 E_\text{cut}^2-Q^2}{2 \left(2 E_\text{cut}-m_b\right)}} \text{d}v{\cdot} Q \, \frac{\text{d}^3 \Gamma^{(n)}}{\text{d} v{\cdot} Q\,\text{d} Q^2\, \text{d}z}  \label{Eq:DoubleRateA} \\
     \frac{\text{d}^2 \Gamma_B^{(n)}}{\text{d} Q^2\, \text{d}z} &=   \int_{\frac{4 E_\text{cut} m_b-m_b^2-4 E_\text{cut}^2-Q^2}{2 \left(2 E_\text{cut}-m_b\right)}}^{\frac{m_b^2+Q^2}{2 m_b}} \text{d}v{\cdot} Q \, \frac{\text{d}^3 \Gamma^{(n)}}{\text{d} v{\cdot} Q\,\text{d} Q^2\, \text{d}z}  \label{Eq:DoubleRateB} \\
          \frac{\text{d}^2 \Gamma_C^{(n)}}{\text{d} Q^2\, \text{d}z} &=   \int_{\frac{4 E_\text{cut} m_b-m_b^2-4 E_\text{cut}^2-Q^2}{2 \left(2 E_\text{cut}-m_b\right)}}^{\frac{mb -2 E_\text{cut} - |z|\sqrt{\left(m_b-2 E_\text{cut}\right){}^2+Q^2 \left(z^2-1\right)}}{1-z^2}} \text{d}v{\cdot} Q \, \frac{\text{d}^3 \Gamma^{(n)}}{\text{d} v{\cdot} Q\,\text{d} Q^2\, \text{d}z}  \label{Eq:DoubleRateC} \,.
\end{align}
\end{subequations}
The difficulty now comes into the game, as we have to take into account the additional constraints on $z$ for regions $B$ and $C$, see Eq.~\eqref{Eq:Regions}. The angular spectrum is obtained by
\begin{subequations}
\begin{align}
  \frac{\text{d} \Gamma_A}{\text{d}z} &= \sum_{n=0}^5 (-1)^n \frac{\text{d}^n}{\text{d}(Q^2)^n}\frac{\text{d}^2 \Gamma_A^{(n)}}{\text{d} Q^2\, \text{d}z} \bigg | _{Q^2=m_c^2} \\
  \frac{\text{d} \Gamma_B}{\text{d}z} &= \sum_{n=0}^5 (-1)^n \frac{\text{d}^n}{\text{d}(Q^2)^n} \left[ \frac{\text{d}^2 \Gamma_B^{(n)}}{\text{d} Q^2\, \text{d}z} \,\theta\left( \frac{-4 E_\text{cut} m_b+m_b^2-Q^2}{m_b^2-Q^2} -z \right) \right]_{Q^2=m_c^2} \\  
  \frac{\text{d} \Gamma_C}{\text{d}z} &= \sum_{n=0}^5 (-1)^n \frac{\text{d}^n}{\text{d}(Q^2)^n} \left[ \frac{\text{d}^2 \Gamma_C^{(n)}}{\text{d} Q^2\, \text{d}z} \,\theta\left( -\frac{-4 E_\text{cut} m_b+m_b^2-Q^2}{m_b^2-Q^2} +z \right) \right]_{Q^2=m_c^2} \,.
\end{align}
\end{subequations}
So we see, that for Regions $B$ and $C$ we get additional delta distribution terms in the variable $z$. Hence after evaluating the $Q^2$ integral with the optical theorem \eqref{Eq:OT}, we can re-sort these contributions according to
\begin{subequations}\label{Eq:AngularSpectrumCut}
\begin{align}
  \frac{\text{d} \Gamma_A}{\text{d}z} &=  \frac{\text{d} \Gamma^{(0)}_A}{\text{d}z} \theta(1 + z)\theta(1 - z) \\
  \frac{\text{d} \Gamma_B}{\text{d}z} &=  \frac{\text{d} \Gamma^{(0)}_B}{\text{d}z} \theta(1 + z)\theta(z_\text{cut} - z) + \sum_{n=1}^5 \frac{\text{d} \Gamma^{(n)}_B}{\text{d}z} \delta^{(n-1)} (z_\text{cut} - z)  \\
  \frac{\text{d} \Gamma_C}{\text{d}z} &=  \frac{\text{d} \Gamma^{(0)}_C}{\text{d}z} \theta(-z_\text{cut} + z) \theta(1 - z) + \sum_{n=1}^5 \frac{\text{d} \Gamma^{(n)}_C}{\text{d}z} \delta^{(n-1)} (z_\text{cut} - z) \,.
\end{align}
\end{subequations}
We obtain the complete differential rate with
\begin{equation}
  \frac{\text{d} \Gamma}{\text{d}z} =  \frac{\text{d} \Gamma_A}{\text{d}z} + \frac{\text{d} \Gamma_B}{\text{d}z}  + \frac{\text{d} \Gamma_C}{\text{d}z} \label{Eq:CompleteRateWithCut}\,.
\end{equation}
The integration to the forward-backward asymmetry or the total rate with the
help of Eq.~\eqref{Eq:TRIntegral} and Eq.~\eqref{Eq:AFBIntegral}, respectively
is now straight-forward. The $\text{d}\Gamma^{(0)}$ pieces need to be integrated
with respect to $z$ in their given limits, while higher order contributions are
fixed by the delta distribution, which we need to treat as usual. Once again, we
have checked our result for the total rate including the cut with previously
calculated results using a different method. For the $A_{FB}$ presented below we
need to remember, that we have imposed the condition $z_\text{cut} \geq 0$. The
cut will produce a non-smooth behaviour at the position of the cut, which we
will investigate later. The 
discussion about the use of moments in the angular variables $z$ is similar to Eq.~\eqref{Eq:Zmoments}, however obstructed due to the cut, which will shift contributions. We will not investigate this further.

\section{Comparison of Expressions to Order $1/m_b^3$}\label{sec:afbanalytic}
First we will examine the analytic expressions from the known observables and the
forward-backward asymmetry up to ${\cal O}(1/m_b^3)$ in the HQE. For an easier comparison of the analytic structure, we expand each result in $\rho = m_c^2/m_b^2$ to order ${\cal O}(\rho^2)$. The full results are given
in Appendix~\ref{sec:appendixA}.
The total rate to this order is then given by
\begin{align}
  \Gamma = \frac{G_F^2 | V_{cb}|^2}{192 \pi^3 m_b^5} \Big[  &\left(1 -8 \rho -12 \rho ^2 \log\rho\right) \left(1  -\frac{\mu_\pi^2}{2 m_b^2} \right) - \frac{\mu_G^2}{2 m_b^2} \Big( 3 -8 \rho  +12 \rho ^2 \log \rho +24 \rho ^2   \Big) \nonumber \\
       &+\frac{\rho_D^3}{6 m_b^3} \Big( 77+48 \log (\rho)-88 \rho +36 \rho ^2 \log  \rho  +24 \rho ^2 \Big)  \nonumber \\
       &+\frac{\rho_{LS}^3}{2 m_b^3} \Big( 3 -8 \rho  +12 \rho ^2 \log \rho +24 \rho ^2  \Big) \Big]\,.\label{Eq:TRmb3Exp}
\end{align}
The moments and forward-backward asymmetry are normalised to the total rate. As we are interested in the dependence on the heavy quark parameters for the fit, we expand the results in $1/m_b$. Note that
starting at order $1/m_b^4$ we encounter mixed terms in this approach, e.g. we have $(\mu_\pi^2)^2$, but to the order we are considering the results, this does not occur. Thus any observable we are considering below, can be viewed as an expansion of a function  given by
\begin{subequations}
\begin{align}
    F & = \frac{\sum_{i=0} n[i]\frac{1}{m_b^i} }{\sum_{j=0} d[j]\frac{1}{m_b^j}} \\
    \Rightarrow\qquad F_\text{exp.} &= \frac{n[0]}{d[0]} + \frac{d[0] n[2]-d[2] n[0]}{d[0]^2 m_b^2}+\frac{d[0] n[3]-d[3] n[0]}{d[0]^2 m_b^3}\,.
\end{align}
\end{subequations}
We have explicitly used the fact, that $1/m_b$ corrections vanish, hence
$n[1]=d[1]=0$. The denominator function $d[i]$ is always given by the total
rate~\eqref{Eq:TRmb3}, while we list the numerator functions $n[i]$ in
Eq.~(\ref{Eq:AFBmb3}-\ref{Eq:MXmb3}). We find for the forward-backward
asymmetry, where we have expanded the result both in $\rho$ and $1/m_b$ for
comparison
\begin{align}
  A_{FB} =  & \frac14 \Big [ 1-12 \rho +12 \rho ^2 \log \rho+ 64 \rho ^{3/2} -186 \rho ^2   \Big] \nonumber \\
    &+ \frac{ 4 \mu_\pi^2}{3 m_b^2} \Big [-1 +6 \sqrt{\rho } -23 \rho  -12 \rho ^2 \log \rho +  68 \rho ^{3/2} -199 \rho ^2   \Big ] \nonumber \\
    &+  \frac{  \mu_G^2}{3 m_b^2}  \Big[-4   +24 \sqrt{\rho }  -92 \rho -48 \rho ^2 \log \rho   + 272 \rho ^{3/2} - 796 \rho ^2 \Big]  \nonumber \\
    &+ \frac{  \rho_D^3}{ 3 m_b^3}  \Big[-14 -6 \log \rho  +24 \rho  \log \rho  +16 \sqrt{\rho } -3 \rho  +1020 \rho ^2 \log \rho   -144 \rho ^2 \log ^2\rho  \nonumber \\
    &\phantom{+}\,+1640 \rho ^2 -384 \rho ^{3/2} \log \rho  -488 \rho ^{3/2}    \Big] +\frac{  \rho_{LS}^3}{ m_b^3}  \Big[-1 -18 \rho ^2 \log \rho -24 \rho ^{3/2}+51 \rho ^2  \Big] \,.\label{Eq:AFBmb3exp}
\end{align}
We see, that especially for the lowest order, there is a similar dependence as for the normalisation and hence the $|V_{cb}|$ extraction. The HQE parameters themselves are 
extracted from moments, currently the charged lepton energy and hadronic invariant mass one. We quote the most important moments~\cite{Mannel:2010wj} in the same way as we have done for the forward-backward asymmetry. The charged lepton energy moment is given by
\begin{align}
  &\langle E_\ell \rangle =  \frac{m_b}{20}\big[ 1 + \frac{\mu _{\pi }^2 }{2
m_b^2} \big] \big[7 - 19 \rho + 96 \rho ^2 \log  \rho  - 272 \rho ^2 \big]
 \nonumber \\
 &-\frac{\mu _G^2}{120 m_b} \big[7288 \rho ^2+695 \rho +48 (67 \rho
+5) \rho \log  \rho +57\big] \nonumber \\
&+\frac{\rho _D^3 }{360
m_b^2}\big[128744 \rho ^2+19008 \rho ^2 \log ^2 \rho +48
\left(2384 \rho ^2+109 \rho +9\right) \log  \rho +8389 \rho
+999\big] \nonumber \\
&+  \frac{\rho _{LS}^3}{40 m_b^2} \big[872 \rho ^2+240 \rho ^2 \log  \rho +17
\rho +3\big]\,.\label{Eq:ELmb3exp}
\end{align}
The partonic invariant mass and energy are related to the hadronic invariant mass by
\begin{equation}
  M_X^2 = (P_B - q)^2 = M_B^2 - 2 M_B v{\cdot} q  + q^2
\end{equation}
and hence we need to introduce the dependence to the mass of the $B$-meson. It is obvious to identify the source of each of the terms below from that equation
\begin{align}
  \langle M_X^2 \rangle & = M_B^2 + m_b M_B \left(\frac{204 \rho ^2}{5}+\frac{72}{5} \rho ^2 \log  \rho +\frac{31 \rho }{10}-\frac{13}{10}\right) \nonumber \\
  &+m_b^2 \left(-\frac{204 \rho ^2}{5}-\frac{72}{5} \rho ^2 \log
    \rho -\frac{21 \rho }{10}+\frac{3}{10}\right) \nonumber \\
  &+\frac{\mu_\pi^2 }{m_b^2} \Big[m_b M_B \left(\frac{102 \rho ^2}{5}+\frac{36}{5} \rho ^2 \log  \rho +\frac{31 \rho }{20}-\frac{13}{20}\right)  \Big] \nonumber \\
  &+\frac{\mu_G^2}{m_b^2} \Big[ m_b M_B \left(\frac{756 \rho ^2}{5}+\frac{312}{5} \rho ^2 \log  \rho +\frac{151 \rho }{12}+4 \rho  \log  \rho +\frac{21}{20}\right) \nonumber \\
  &+m_b^2 \left(-\frac{2098 \rho
   ^2}{15}-\frac{292}{5} \rho ^2 \log  \rho -\frac{34 \rho }{3}-4 \rho  \log  \rho -\frac{4}{5}\right)\Big] \nonumber \\
  &+\frac{\rho_D^3}{m_b^3} m_b^2 \Big[\frac{82}{5} ++\frac{5026 \rho }{45}+ \frac{12790 \rho ^2}{9}   +\frac{964}{15} \rho  \log (\rho
   )+\frac{28 \log  \rho }{5}  \nonumber \\
   &+\frac{912}{5} \rho ^2 \log ^2 \rho +\frac{3592}{3} \rho ^2 \log  \rho \Big] \nonumber \\
  &+\frac{\rho_{LS}^3}{m_b^3} \Big[m_b^2 \left(\frac{2098 \rho ^2}{15}+\frac{292}{5} \rho ^2 \log  \rho +\frac{34 \rho }{3}+4 \rho  \log  \rho +\frac{4}{5}\right) \Big] \,.\label{Eq:MX2mb3exp}
\end{align}
For an easier comparison of the functional form of the expanded results, we
quote the dependence numerically using the numerical result from
~\cite{Alberti:2014yda} with $\rho \approx 0.047 $. Here we have only expanded
in $1/m_b$ and keep the full dependence on $\rho$
\begin{align}
  \Gamma &\approx     0.706-\frac{0.353 \mu _{\pi }^2}{m_b^2} -\frac{1.297 \mu _G^2}{m_b^2} -\frac{12.350 \rho _D^3}{m_b^3} +\frac{1.297 \rho _{\text{LS}}^3}{m_b^3}\\
     A_{FB} &\approx      0.135-\frac{0.376 \mu_{\pi
}^2}{m_b^2}-\frac{1.197\mu_G^2}{m_b^2}-\frac{0.570 \rho _D^3}{m_b^3}
-\frac{0.516 \rho _{\text{LS}}^3}{m_b^3} \label{Eq:AFBmb3expnum}
\end{align}
\begin{align}
    \langle E_\ell \rangle &\approx m_b \left[ 0.316+\frac{0.158 \mu _{\pi }^2}{m_b^2} -\frac{0.379 \mu _G^2}{m_b^2} -\frac{1.999 \rho _D^3}{m_b^3}+\frac{0.087\rho _{\text{LS}}^3}{m_b^3}  \right] \\
    \langle m_X^2 \rangle &\approx \frac{1}{m_b} \bigg[M_B^2 m_b-1.187 M_B m_b^2+0.234 m_b^3 -0.594 \mu _{\pi }^2 M_B    \nonumber \\
		      &\phantom{\approx}\,+\mu _G^2 \left(0.890 M_B -0.590 m_b\right) -5.471 \rho _D^3+0.590 \rho _{\text{LS}}^3   \bigg] \,.
\end{align}
From this, we can see that the coefficients of $\mu_\pi^2$ and $\mu_G^2$ have opposite signs for the moments, while same sign coefficients for the rate and the forward-backward asymmetry. It has been known before, that the sensitivity to $\mu_G^2$ and $\rho_{LS}^3$ is low for all currently used observables. The sensitivity to $\mu_G^2$ is enhanced for $A_{FB}$ and therefore we gain useful information. Furthermore the higher order contributions
seem to be stronger suppressed for the $A_{FB}$. Hence we are able to extract a further linear combination, which is especially useful for the normalisation. In that sense, the value of $\mu_G^2$ seems to be stronger constraint.

\section{Numerical Results to Order $1/m_b^5$}\label{sec:afbnumerical}

\begin{figure}[htp]
 \begin{center}
 \includegraphics[scale=0.35]{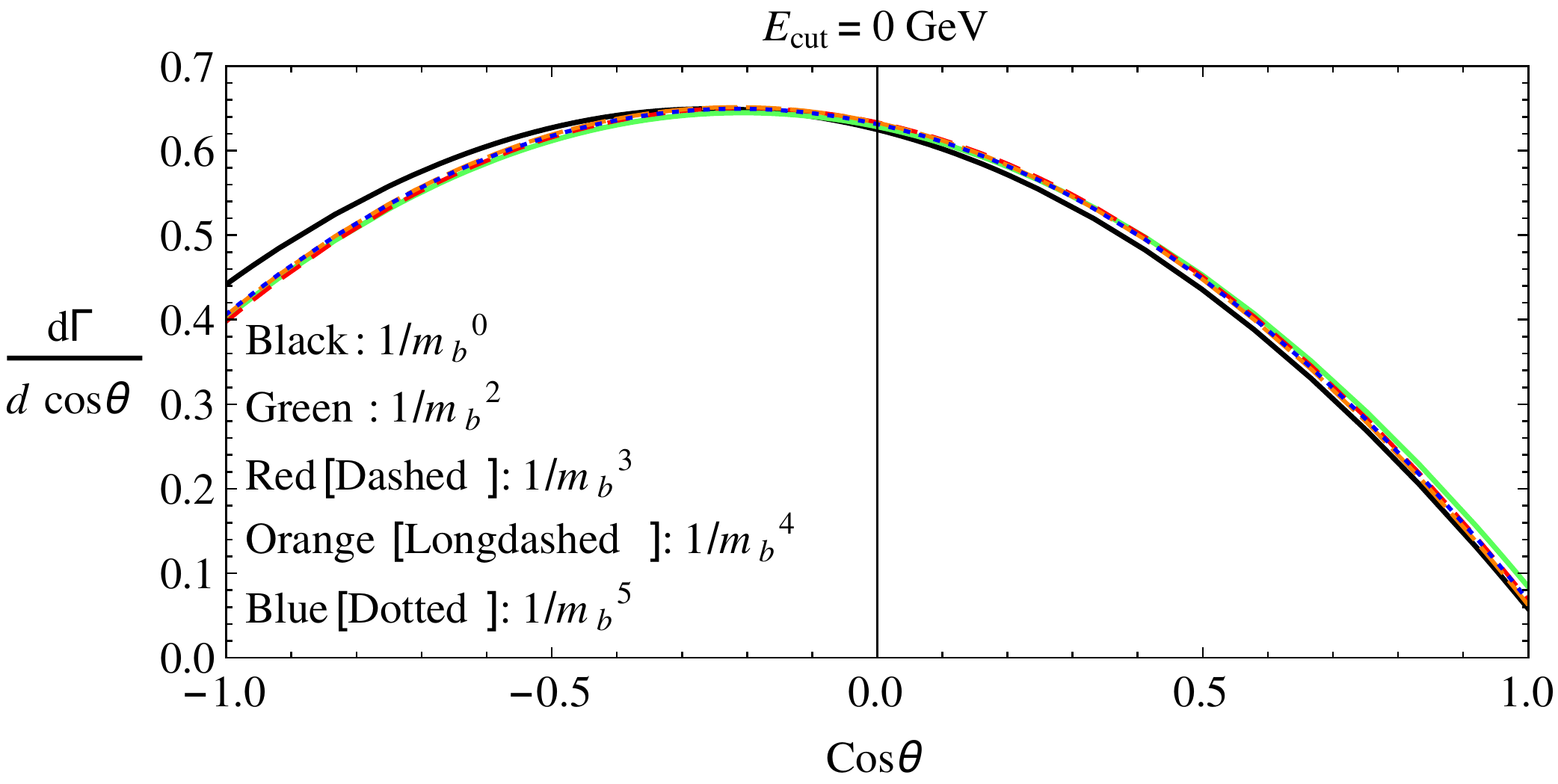}\hfill
\includegraphics[scale=0.35]{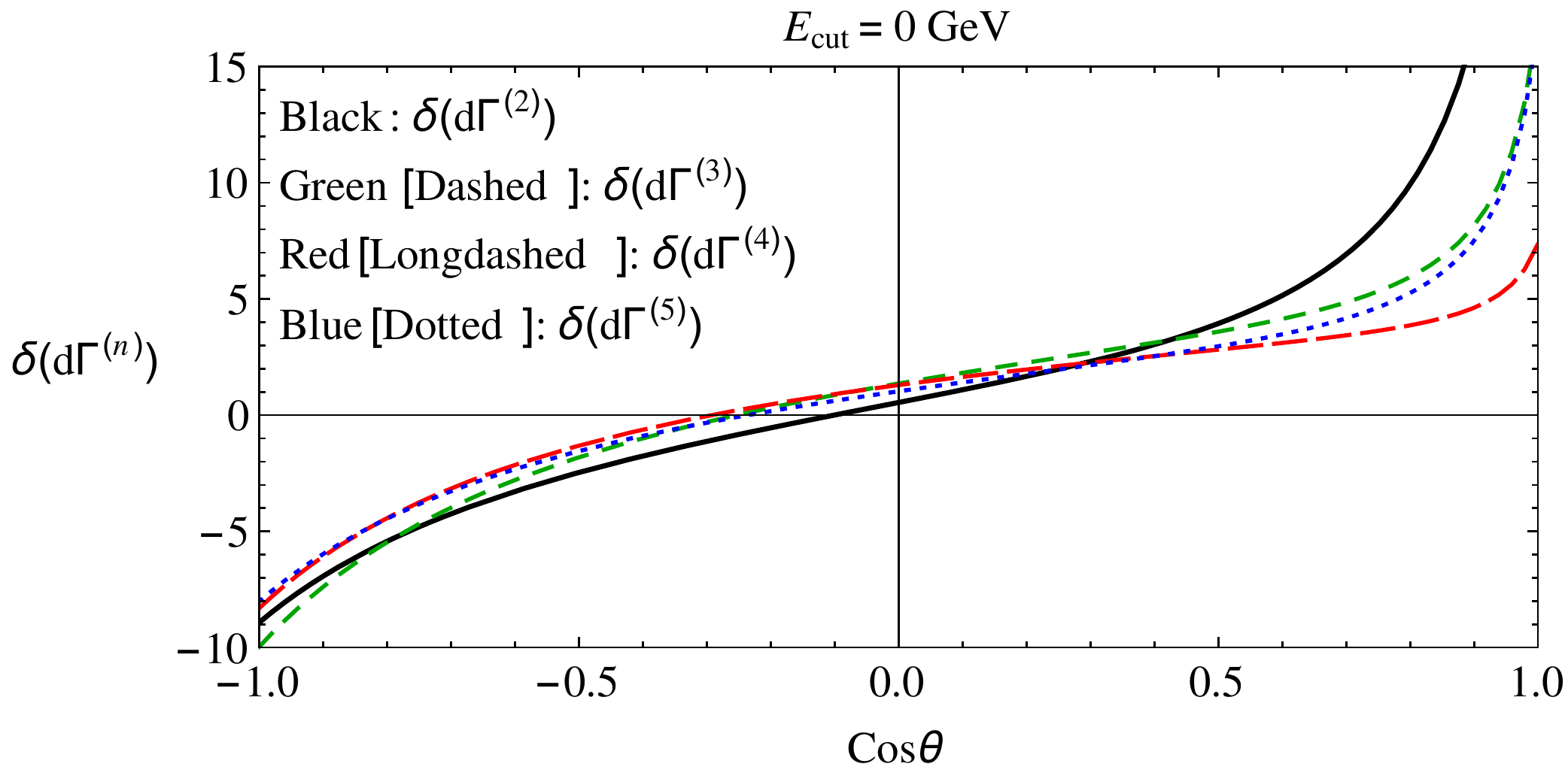}
\end{center}
\caption{The differential rate $\frac{\text{d}\Gamma}{\text{d}\cos\theta}$ as a
function of the angle $z=\cos \theta$ with no energy cut on the charged lepton.
Left: The spectrum itself for various orders in $1/m_b$. Right: Relative
contribution $\delta =
100 \,\frac{\text{d}\Gamma^{(n)}-\text{d}\Gamma^{(3)}}{\text{d}\Gamma^{(3)}}$
from
order $1/m_b^n$ to the partonic rate.}\label{Fig:dGammadzCut0}
\end{figure}

First we investigate the differential spectrum in $z= \cos \theta$ itself. On the left-hand side in
Fig.~\ref{Fig:dGammadzCut0} we have displayed the
spectrum itself with no minimum energy cut on the charged lepton. The
individual colour coded curves are contributions including $1/m_b^n$ corrections
to the order: $1/m_b^0$ (black), $1/m_b^2$ (green), $1/m_b^3$ (red dashed),
$1/m_b^4$ (orange long-dashed) and $1/m_b^5$ (blue dotted).   

We find, that the corrections are getting larger for approaching the physical
endpoints of the angle. As the rate is approaching zero for $z=\cos \theta
\rightarrow 1$, the absolute deviations are hardly visible in the left plot,
while the corrections for $z \rightarrow -1$ are larger in absolute values and
hence visible in the plot, although the relative corrections are smaller.
Roughly, the corrections for negative $z$ are negative, while for
positive $z$ they are positive. These interesting facts seem to be related to
the kinematics of the final state charm system, which is sensitive to heavy
quark corrections. The corrections themselves are in reasonable size and behave
as expected for higher-orders. For this remember, that the hadronic tensor
depends on $v{\cdot} q$ and $q^2$, while $z$ is a function of $v{\cdot} q$ and
$q^2$, as well as the charged electron energy.

From the left plot, we can see the asymmetric behaviour of the spectrum, and
hence a forward-backward asymmetry can be observed. Especially we find for
higher-order corrections, that the $1/m_b^2$ corrections are very important.
The convergence of higher order terms in the expansion is good and stable. Hence
in combination with the fact, that we are sensitive to a particular combination
of $\mu_\pi^2$ and especially $\mu_G^2$, see Eqs.~\eqref{Eq:AFBmb3exp}
and~\eqref{Eq:AFBmb3expnum}, from this observable we have a good sensitivity to
$\mu_G^2$.

\begin{figure}[htp]
 \begin{center}
 \includegraphics[scale=0.35]{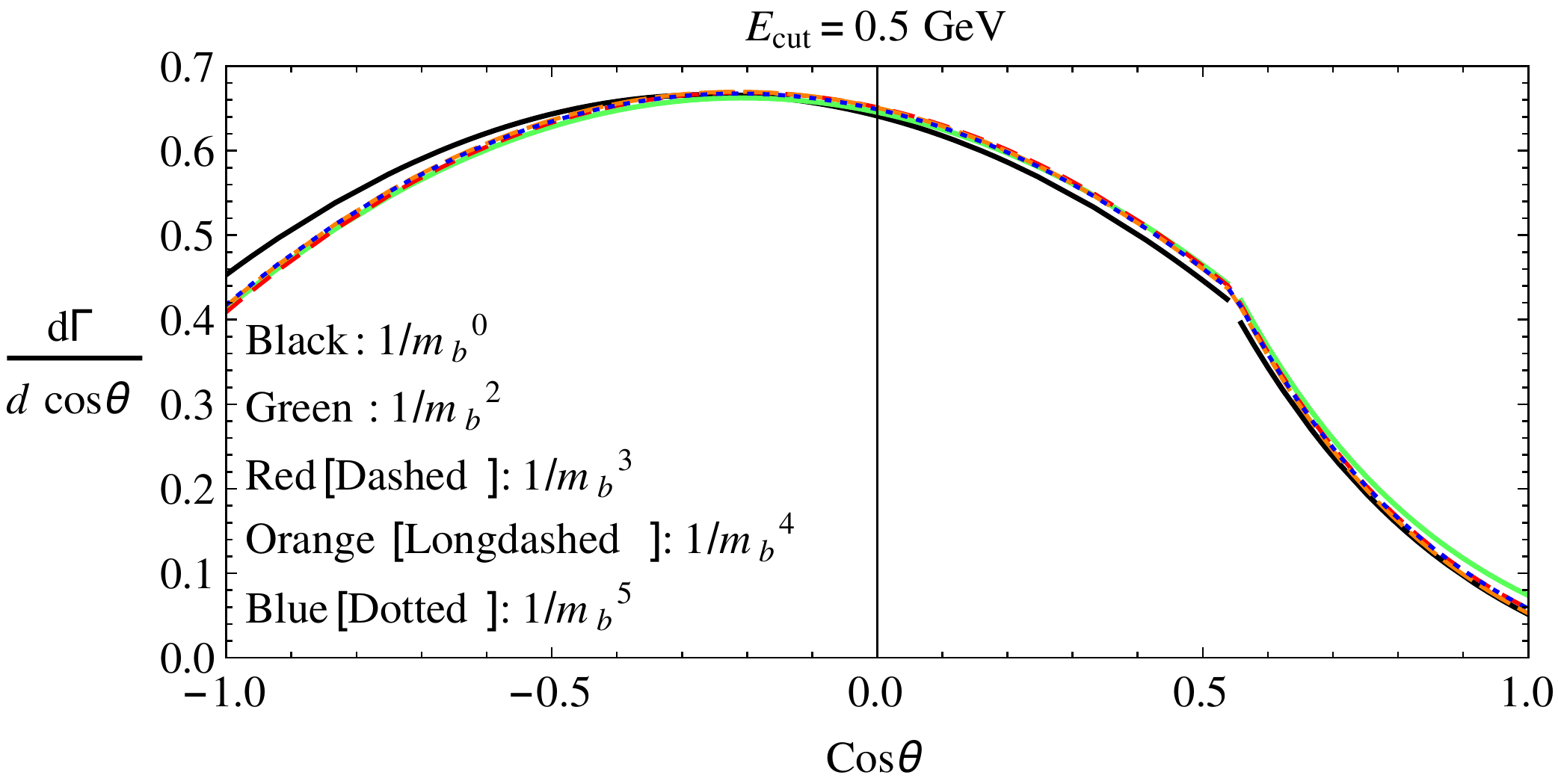}\hfill
\includegraphics[scale=0.35]{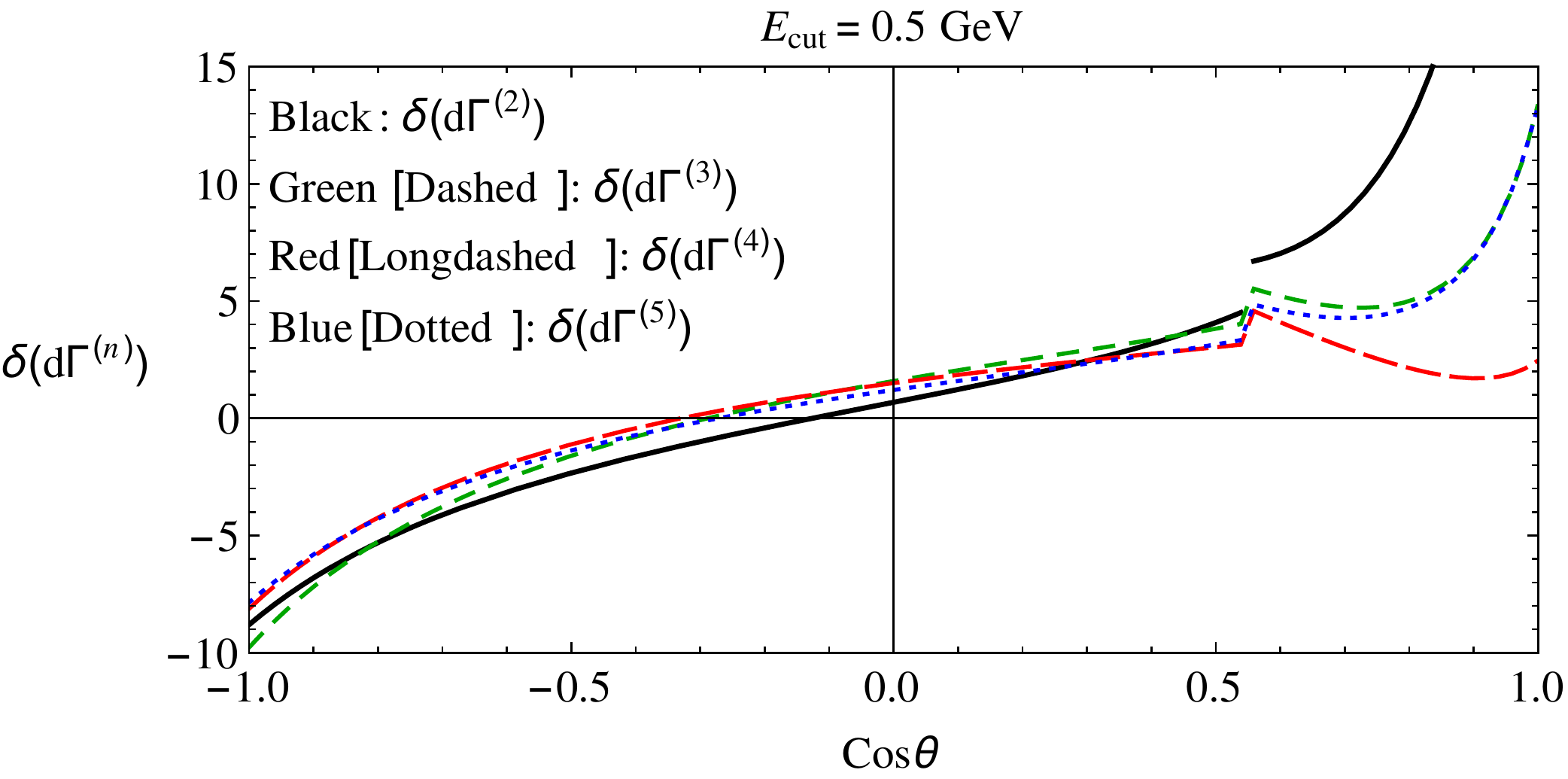}
\end{center}
\caption{The differential rate $\frac{\text{d}\Gamma}{\text{d}\cos\theta}$ as a
function of the angle $z=\cos \theta$ with a minimum energy cut on the charged
lepton of $E_\text{cut} = 0.5 \text{ GeV}$.
Left: The spectrum itself for various orders in $1/m_b$, right relative
contribution $\delta =
100 \,\frac{\text{d}\Gamma^{(n)}-\text{d}\Gamma^{(3)}}{\text{d}\Gamma^{(3)}}$
from
order $1/m_b^n$ to the partonic rate.}\label{Fig:dGammadzCut05}
\end{figure}

In real experimental environments, we have to impose a minimum cut on the
charged lepton energy. In the following we investigate the consequences on the
differential spectrum. A realistic cut from current experiments is
$E_\text{cut} = 1\text{ GeV}$ \cite{Alberti:2014yda}, while the hope is to
reduce this in future experiments to even lower values. It is well-known that
restricting the
phase-space limits the validity of the heavy quark expansion and hence
higher-order terms have a larger impact. Currently it is estimated, that the HQE
works
still to a reasonable precision for $E_\text{cut} \lesssim 1.5\text{ GeV}$
\cite{Mannel:2010wj}, but $E_\text{cut} \approx 1\text{ GeV}$ is certainly
preferred.

\begin{figure}[hbp]
 \begin{center}
 \includegraphics[scale=0.35]{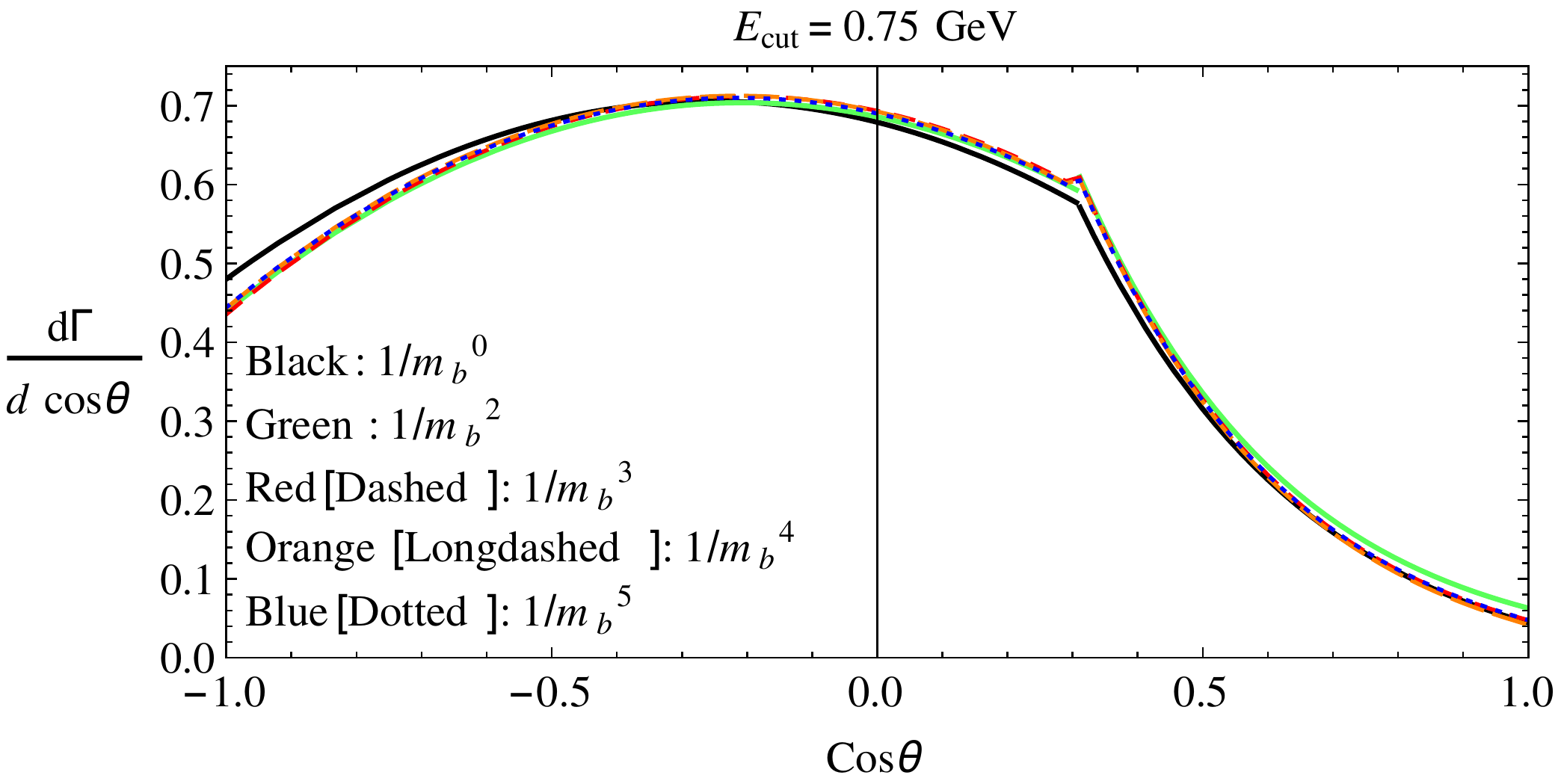}\hfill
\includegraphics[scale=0.35]{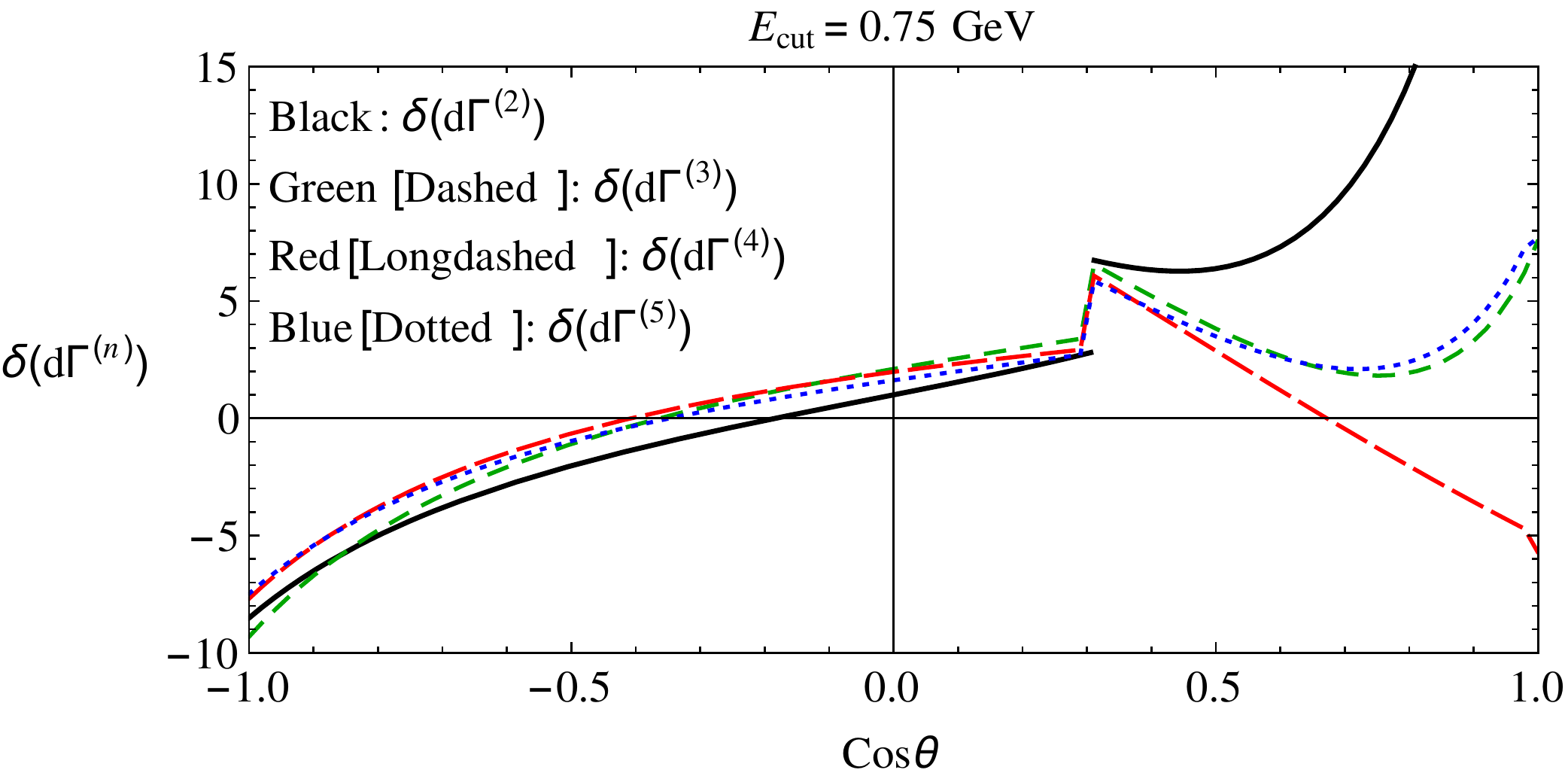}
\end{center}
\caption{The differential rate $\frac{\text{d}\Gamma}{\text{d}\cos\theta}$ as a
function of the angle $z=\cos \theta$ with a minimum energy cut on the charged
lepton of $E_\text{cut} = 0.75 \text{ GeV}$.
Left: The spectrum itself for various orders in $1/m_b$, right relative
contribution $\delta =
100 \,\frac{\text{d}\Gamma^{(n)}-\text{d}\Gamma^{(3)}}{\text{d}\Gamma^{(3)}}$
from
order $1/m_b^n$ to the partonic rate.}\label{Fig:dGammadzCut075}
\end{figure}

In Fig.~\ref{Fig:dGammadzCut05} we compare the same plots as before, now
imposing a cut of $E_\text{cut} = 0.5\text{ GeV}$.  In this scenario, we
observe a kink in the theoretical spectrum exactly at the cut separation
\begin{equation}
  z_\text{cut} = 1-\frac{4 E_\text{cut} m_b}{m_b^2-m_c^2}\,.
\end{equation}
We find, that for the partonic rate the spectrum behaves unsteady at this
position. For higher-orders this becomes worse and we find a discontinuity.
This exactly reflects the fact, that we are expanding a non-local object into
local terms. In reality this kink will be smoothed out by the distribution of
the final state mass. 

\begin{figure}[htp]
 \begin{center}
 \includegraphics[scale=0.35]{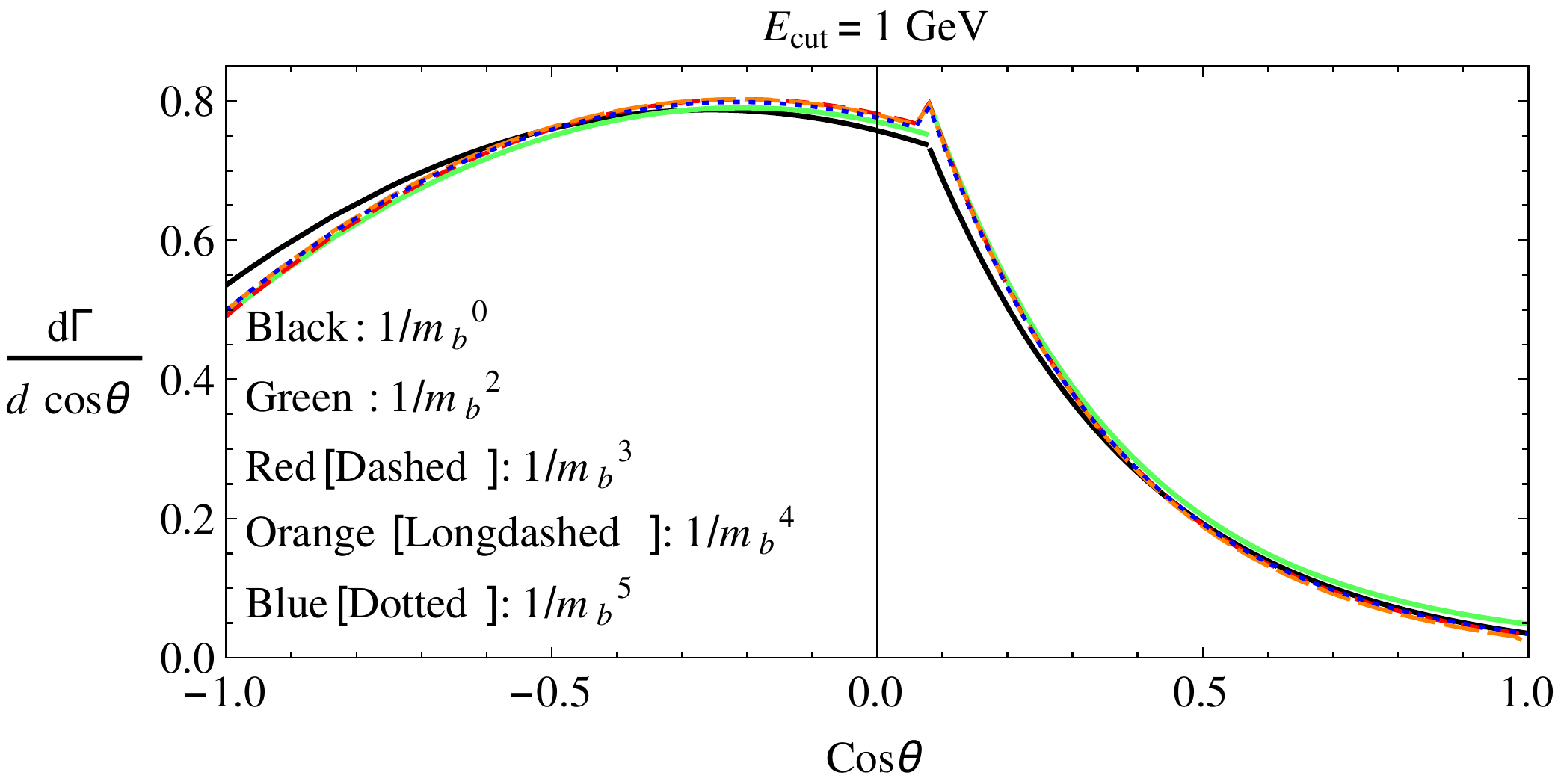}\hfill
\includegraphics[scale=0.35]{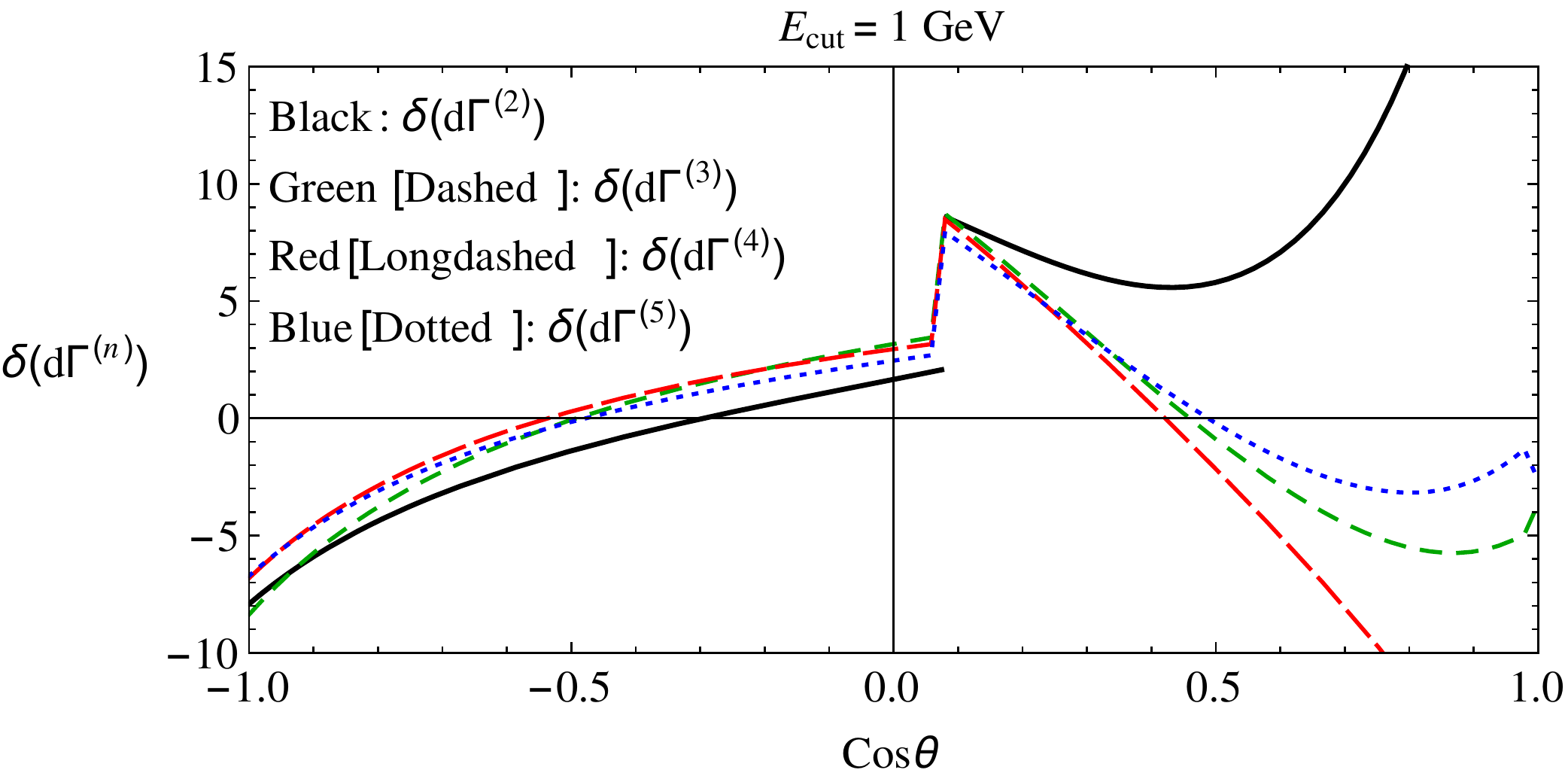}
\end{center}
\caption{The differential rate $\frac{\text{d}\Gamma}{\text{d}\cos\theta}$ as a
function of the angle $z=\cos \theta$ with a minimum energy cut on the charged
lepton of $E_\text{cut} = 1 \text{ GeV}$.
Left: The spectrum itself for various orders in $1/m_b$, right relative
contribution $\delta =
100 \,\frac{\text{d}\Gamma^{(n)}-\text{d}\Gamma^{(3)}}{\text{d}\Gamma^{(3)}}$
from
order $1/m_b^n$ to the partonic rate.}\label{Fig:dGammadzCut1}
\end{figure}

As obvious from Figs.~\ref{Fig:dGammadzCut075} and \ref{Fig:dGammadzCut1}, the
latter has a cut of $E_\text{cut} = 1 \text{ GeV}$ used in current data, the
cut moves to smaller values of $z$ and the discontinuity is enhanced. While the
relative corrections in the right side plot are stable for $z < z_\text{cut}$ 
they are getting larger for $z > z_\text{cut}$ and the hierarchy of
corrections to various orders is clearly visible. Interestingly the effect of
${\cal O}(1/m_b^4)$ seems to be stronger, while ${\cal O}(1/m_b^5)$ approaches
${\cal O}(1/m_b^3)$ and both of the latter seem to be more stable for
$z\rightarrow 1$. Please note, that the spectrum is shifted towards the negative
values of $z$ with increasing cut.

\begin{figure}[htp]
 \begin{center}
 \includegraphics[scale=0.35]{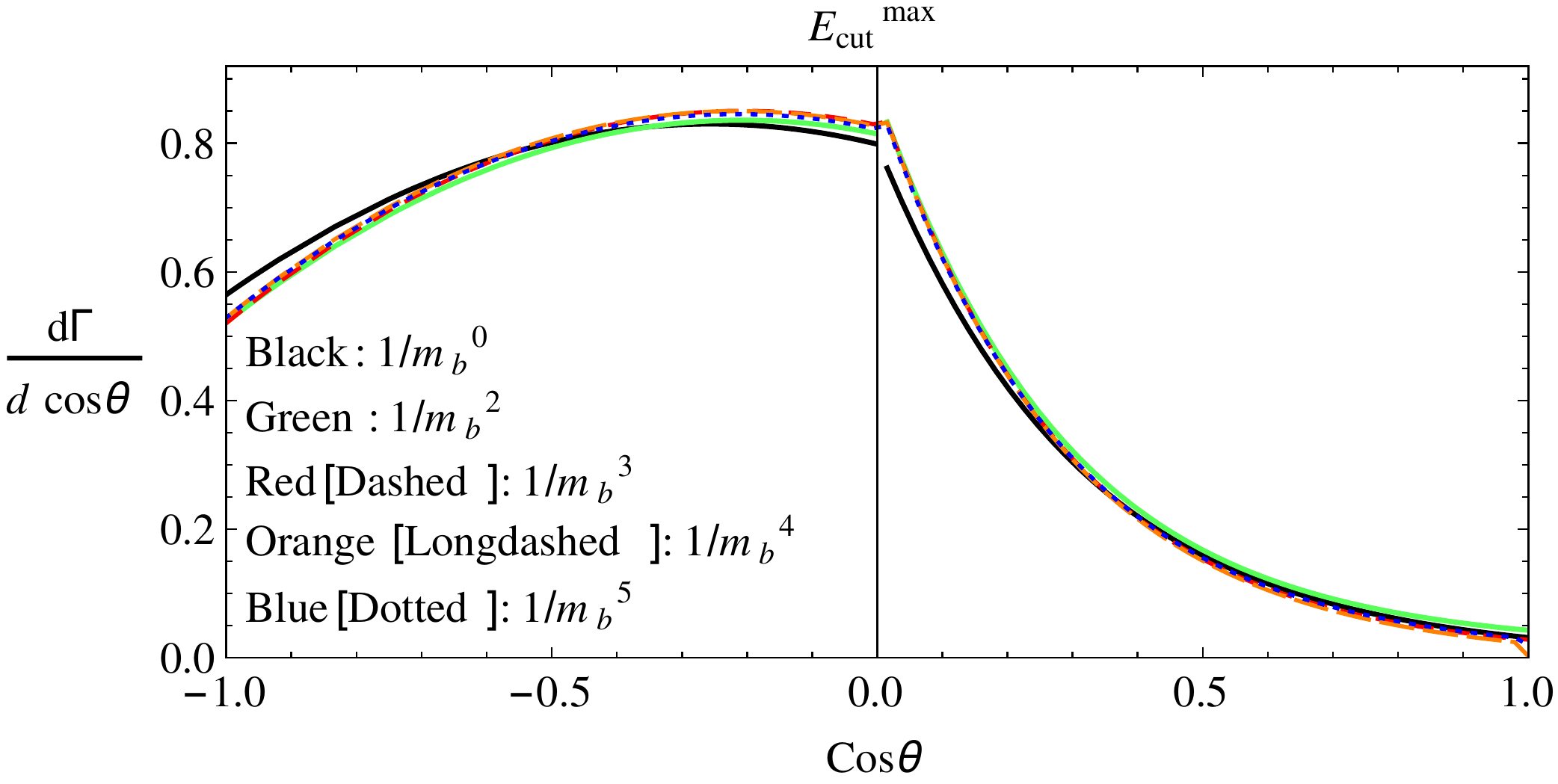}\hfill
\includegraphics[scale=0.35]{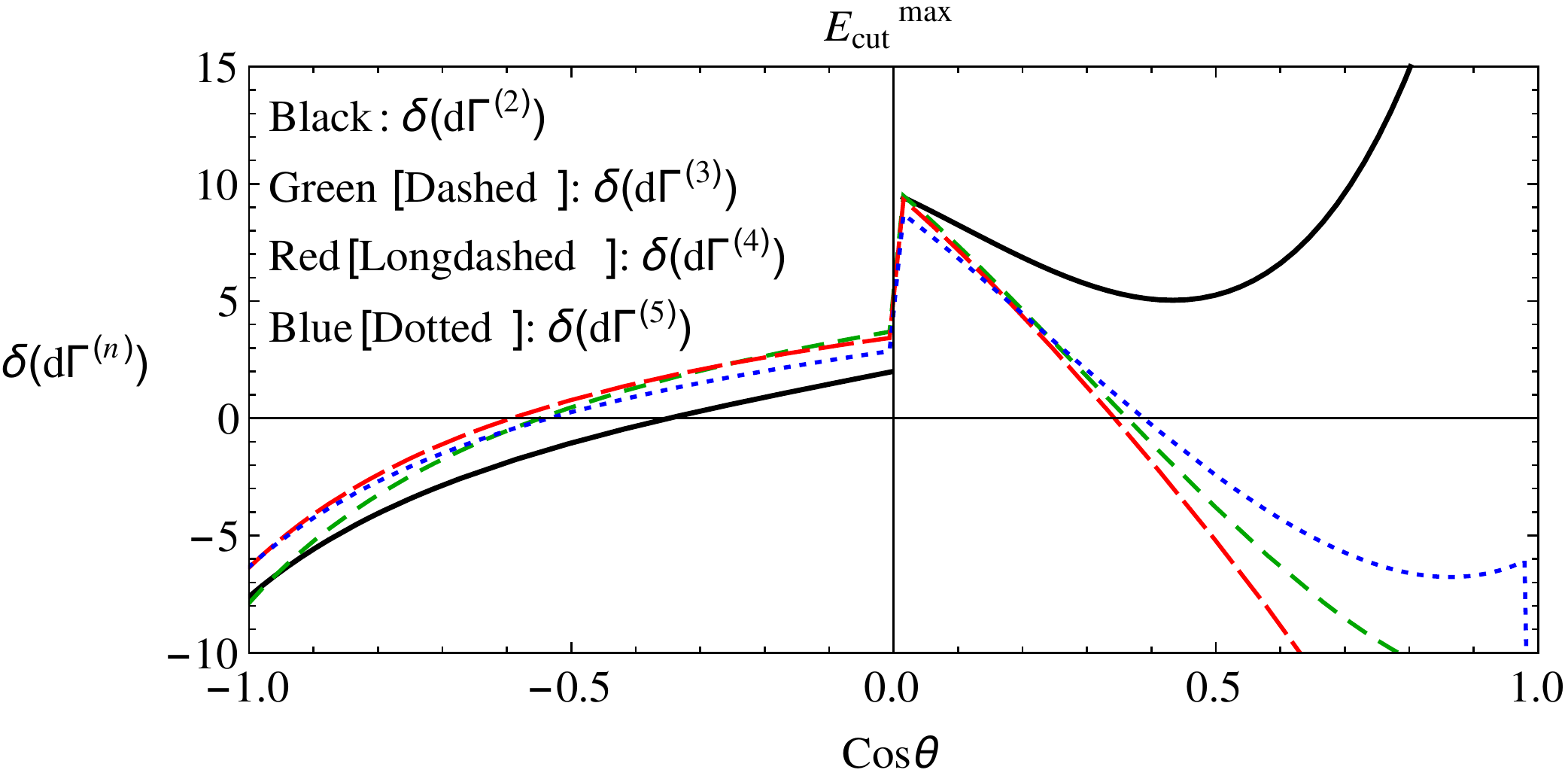}
\end{center}
\caption{The differential rate $\frac{\text{d}\Gamma}{\text{d}\cos\theta}$ as a
function of the angle $z=\cos \theta$ with a maximal energy cut on the charged
lepton of $E_\text{cut} = \frac{m_b^2-m_c^2}{4 m_b} \approx 1.08 \text{ GeV}$,
such that the cut separation is still positive.
Left: The spectrum itself for various orders in $1/m_b$, right relative
contribution $\delta =
100 \,\frac{\text{d}\Gamma^{(n)}-\text{d}\Gamma^{(3)}}{\text{d}\Gamma^{(3)}}$
from
order $1/m_b^n$ to the partonic rate.}\label{Fig:dGammadzCutmax}
\end{figure}

For the maximal cut of  $E_\text{cut}^\text{max} = \frac{m_b^2-m_c^2}{4 m_b}$ we
find in Fig.~\ref{Fig:dGammadzCutmax}, that the separation is exactly at $z=0$,
which was our definition for the maximal allowed cut in this scenario. Of course
theoretically the cut may even be larger, but then our predictions for the
$A_{FB}$ would have to be modified, as terms shift from positive to negative.
As stated before on the one hand this maximal cut is above the current
experimentally used cuts, and on the other hand the larger the cut the less
precise are our predictions and hence our restriction.

In general, only fully integrated observables over the hadronic kinematics are
investigated~\cite{Mannel:2010wj}, with the only exception of the distribution
in the charged lepton energy. The heavy-quark spin-symmetry is only valid for
fully integrated observables over the hadronic part, as it starts from a
spherical symmetry. The charged lepton energy is (mainly) independent from the
hadronic kinematics and hence can be utilised as an additional observable. Here
we find this feature, that $z$ strongly depends on the hadronic kinematics,
which is reflect by this unsteady behaviour.

For this particular observable we find, that the correction
seem to
play a more important role for
$z \rightarrow 1$. That indicates the relation to the final state kinematics of
the hadron system. In
exclusive transitions HQE works fine, if the final and initial state hadron is
moving with the same velocity, while it breaks down for a vastly different
situation. That effect seems to be resembled in this particular spectrum,
although we are investigating a property of the leptonic system, its kinematics
is connected to the hadron system.

We will comment more on the situation and use of this spectrum in
Sec.~\ref{sec:discussion}, and turn now to the integrated forward-backward
asymmetry $A_{FB}$.

\begin{figure}[hbp]
 \begin{center}
 \includegraphics[scale=0.335]{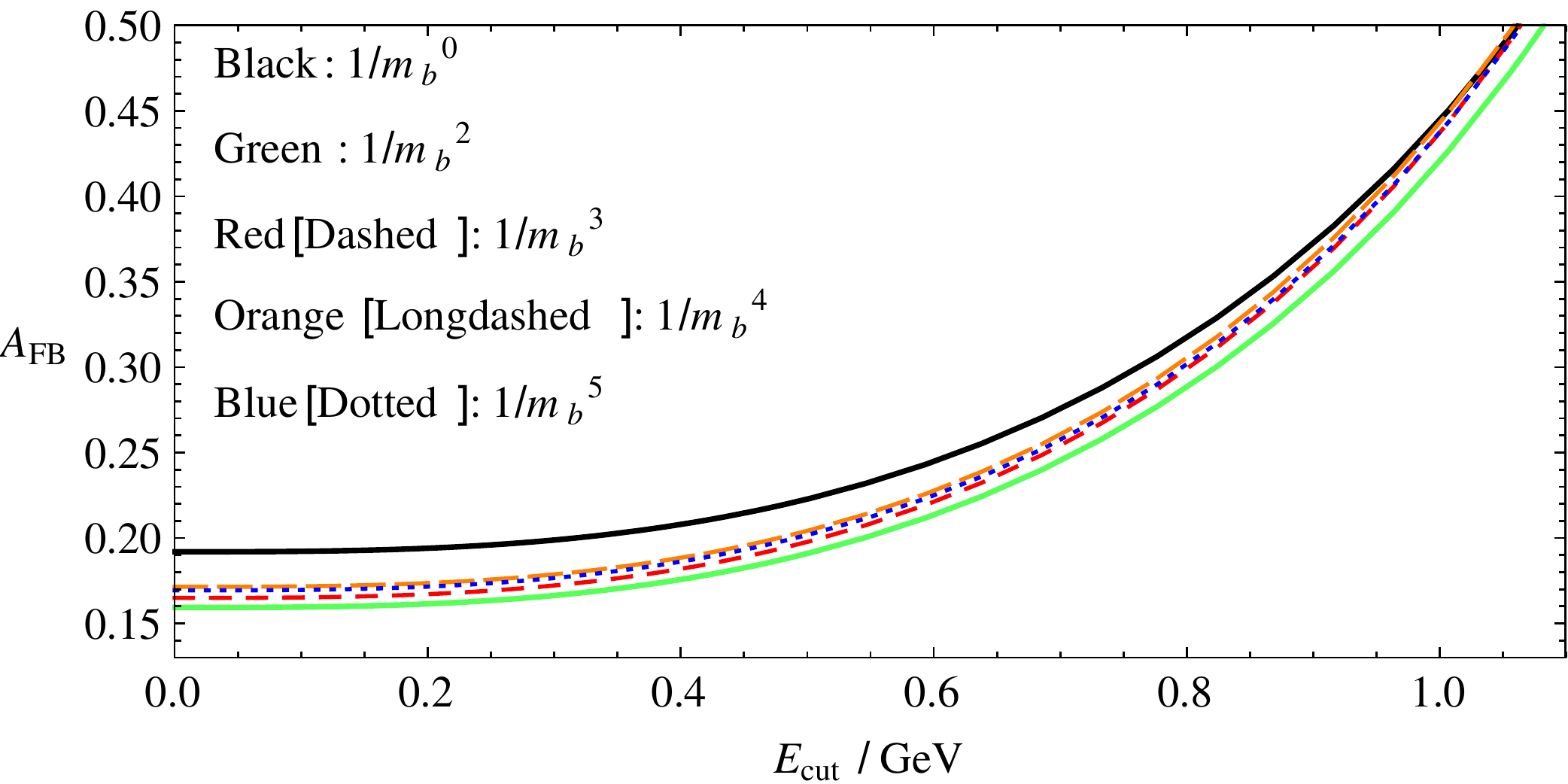}\hfill
\includegraphics[scale=0.365]{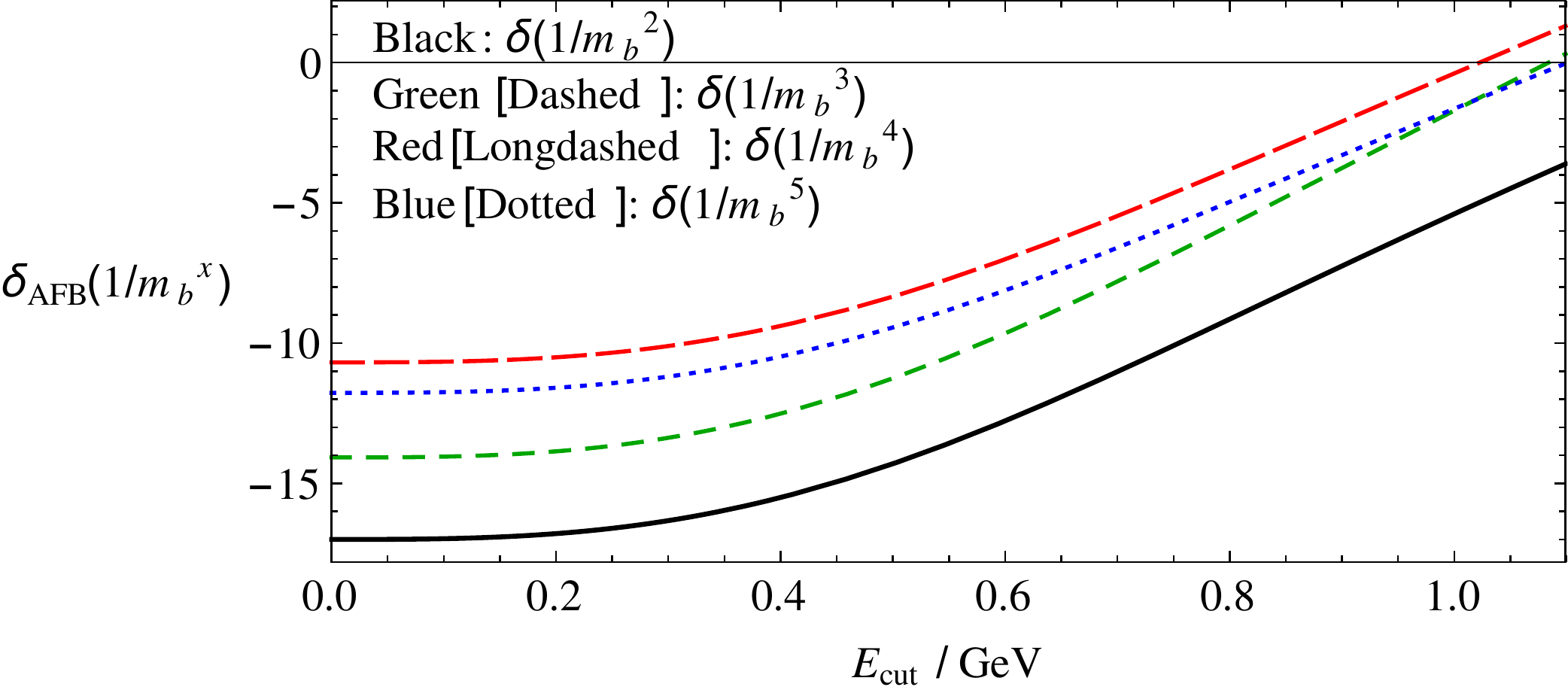}
\end{center}
\caption{The forward-backward asymmetry $A_{FB}$ as a function
of the minimal energy cut on the charged lepton. Left:
$A_{FB}$ for various orders in $1/m_b$, right relative
contribution $\delta = 100 \,\frac{A_{FB}^{(n)}-A_{FB}^{(3)}}{A_{FB}^{(3)}}$
from
order $1/m_b^n$ to the partonic contribution.}\label{Fig:CutDependence}
\end{figure}

We have plotted the forward-backward
asymmetry $A_{FB}$ in Fig.~\ref{Fig:CutDependence} as a function of the cut in
the left plot. The colour coding is: contributions included up to $1/m_b^0$
(black), $1/m_b^2$ (green), $1/m_b^3$ (red dashed), $1/m_b^4$
(orange long-dashed) and $1/m_b^5$ (blue dotted). In the same way as before, we
plot the relative correction 
\begin{equation}
  \delta = 100 \,\frac{A_{FB}^{(n)}-A_{FB}^{(3)}}{A_{FB}^{(3)}}\,
\end{equation}
on the right hand side of the figure. The corrections of $1/m_b^2$ are by far
the biggest one, and hence we are sensitive to them. As obvious from the
spectrum, for a larger cut on the charged lepton energy, we find an increasing
$A_{FB}$. The relative corrections are, contrary to naive expectations,
decreasing for a higher cut. That effect is most probably driven by the fact, that the $A_{FB}$ increases for larger $E_\text{cut}$, and the partonic contribution is growing obviously faster. Even though we have an increasing
forward-backward asymmetry with larger cuts, the absolute difference of the
higher order terms  is larger for a smaller cut. The effect of including higher
order is as expected getting smaller, however again we find that the pure
$1/m_b^5$ corrections have the opposite sign, as can be seen from the right plot
in Fig.~\ref{Fig:CutDependence}.

We may now investigate the stability of this sensitivity to the $1/m_b^2$
parameters while including higher-order terms along the line
of~\cite{Mannel:2010wj}. Defining an observable as $\mathcal{M}^{(n)}$, where
$n$ denotes the order in $1/m_b^n$, we can assess the effect to a single
heavy-quark parameter (HQP) with including higher-order terms by
\begin{equation}
  \delta \text{HQP} = - \frac{\mathcal{M}^{(5)}-\mathcal{M}^{(3)}
}{\frac{\partial \mathcal{M}^{(3)}}{\partial \text{HQP}}}\,. \label{Eq:DeltaHQP}
\end{equation}

\begin{figure}[htp]
 \begin{center}
 \includegraphics[scale=0.35]{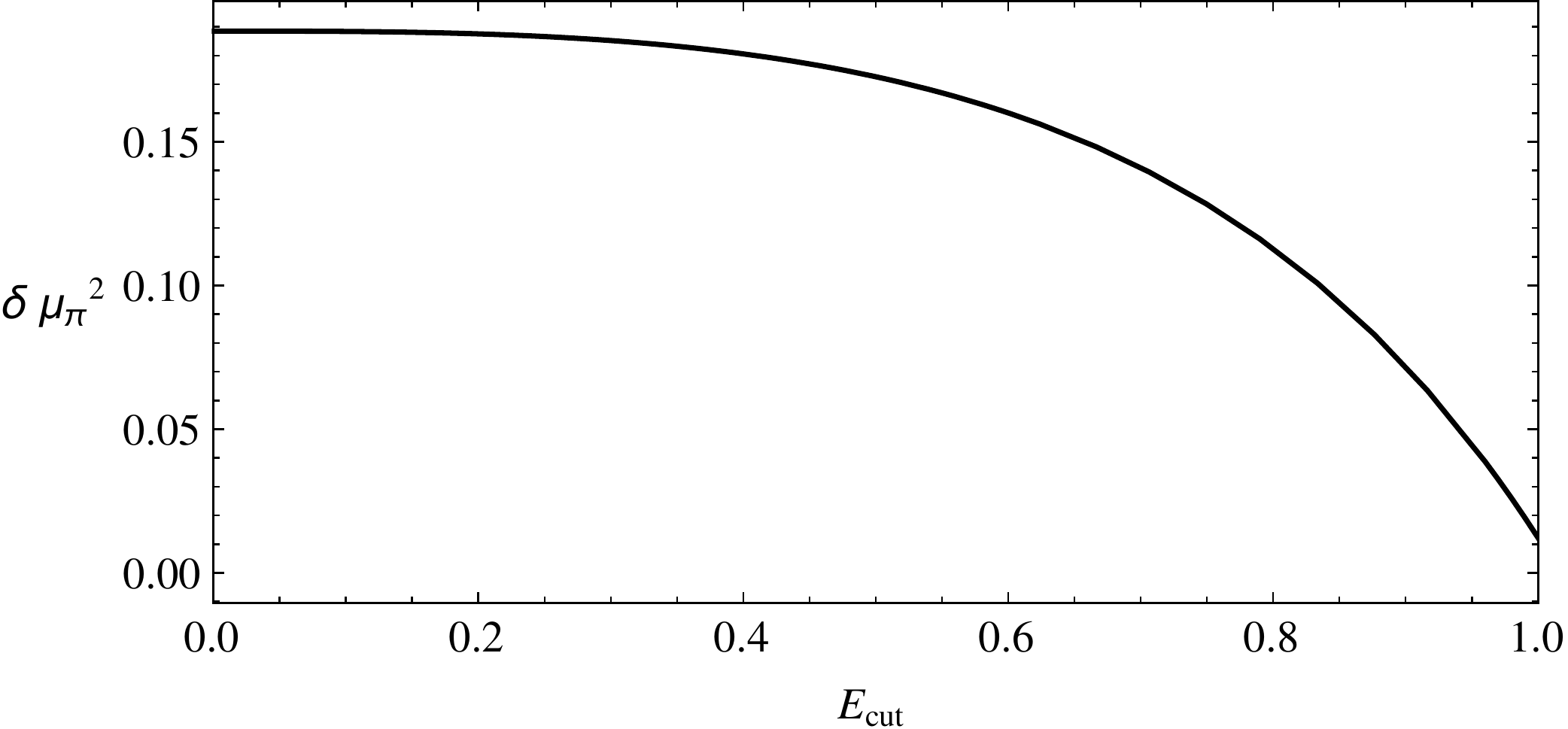}\hfill
\includegraphics[scale=0.35]{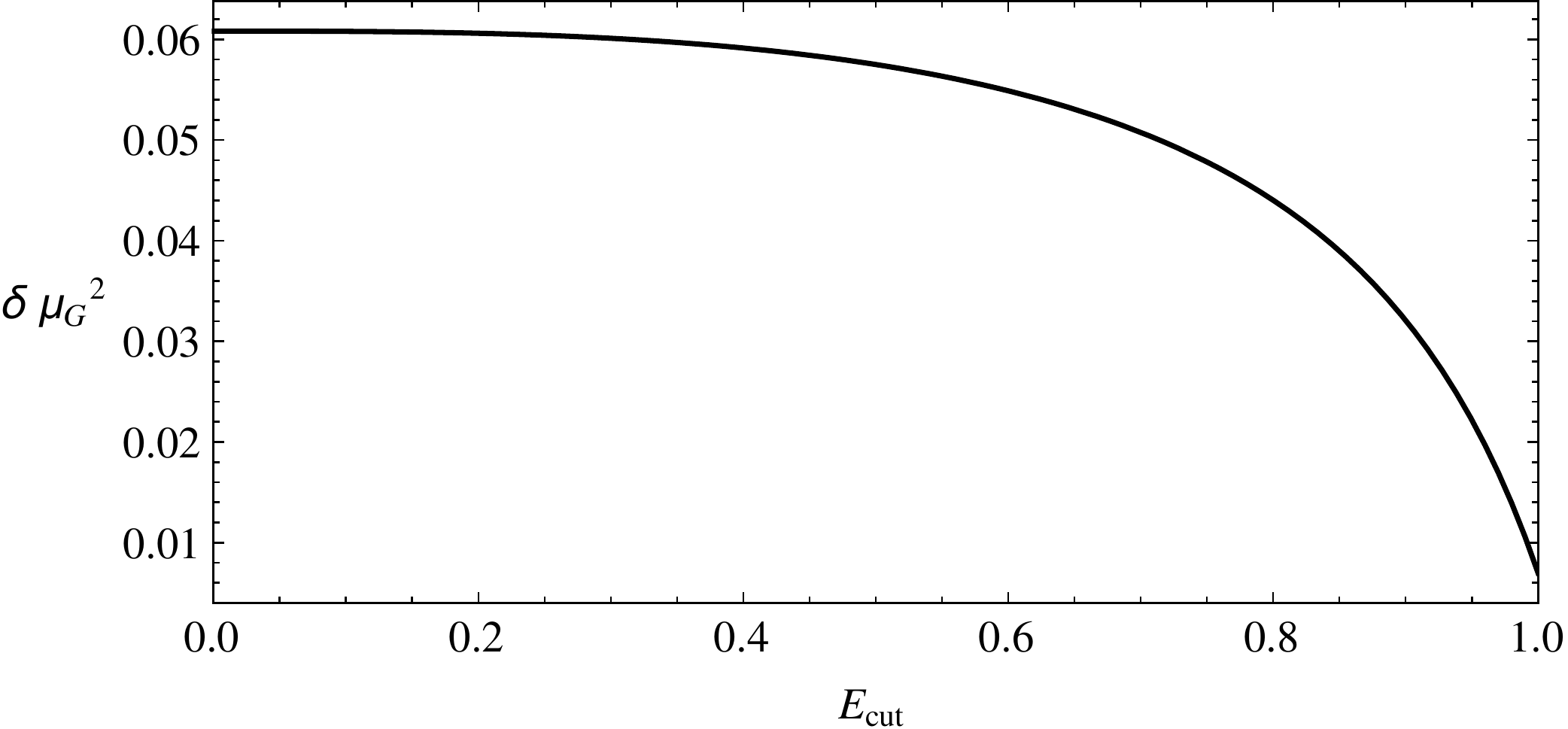}
\end{center}
\caption{Estimating the effect on $\mu_\pi^2$ and $\mu_G^2$ of including
$1/m_b^{4,5}$ in the forward-backward asymmetry using Eq.~\eqref{Eq:DeltaHQP}
as a function of $E_\text{cut}$. }\label{Fig:MUPIMUGdependence}
\end{figure}

\begin{figure}[htp]
 \begin{center}
 \includegraphics[scale=0.35]{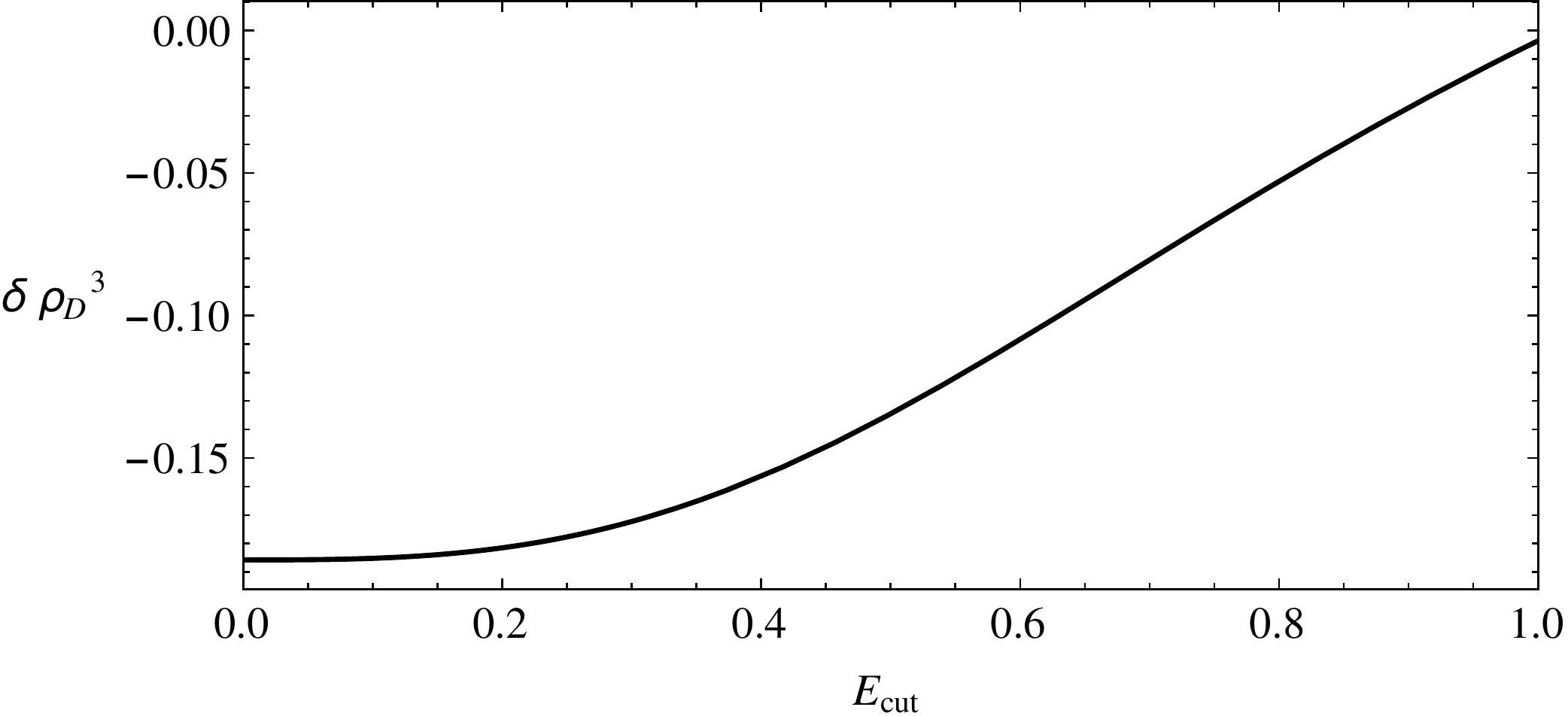}\hfill
\includegraphics[scale=0.35]{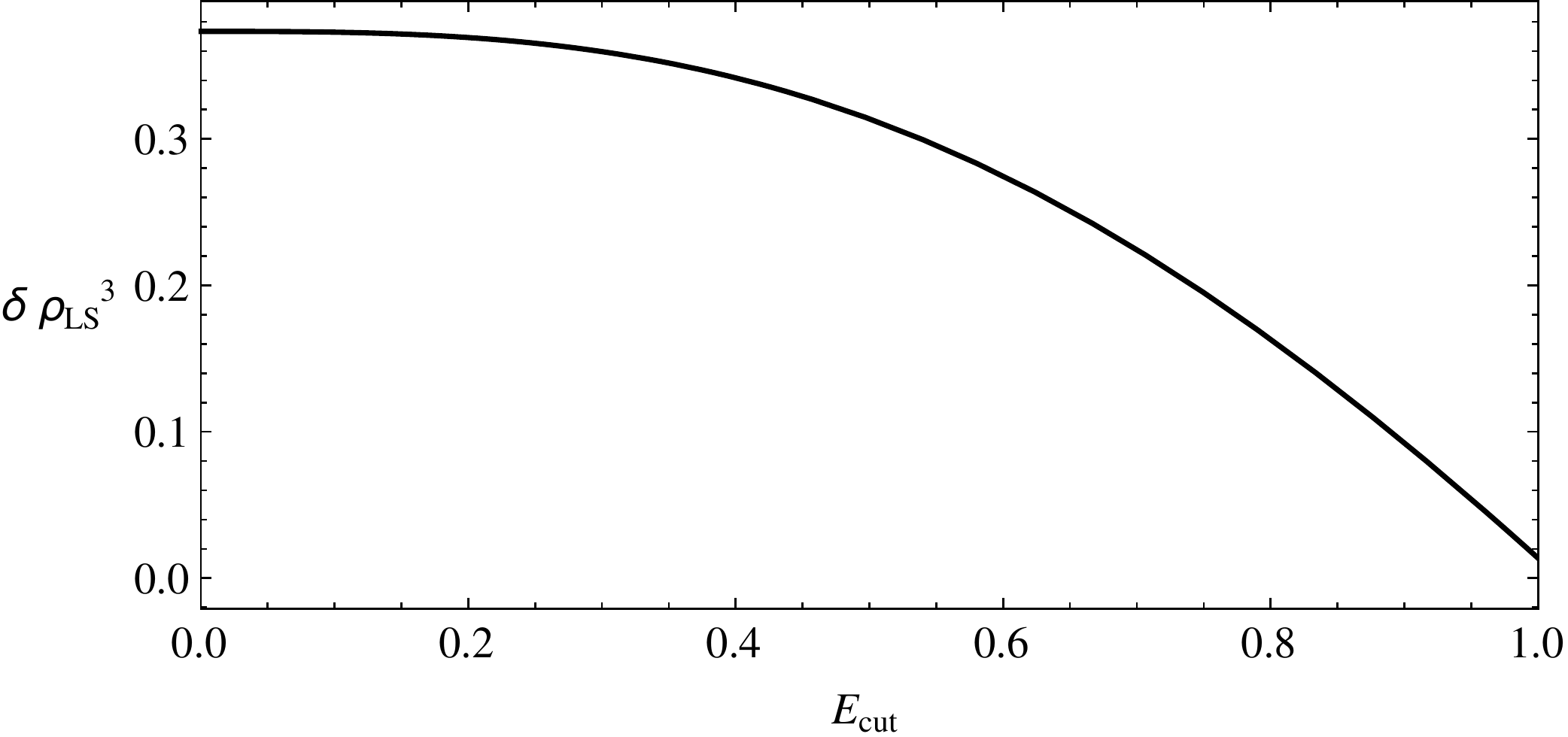}
\end{center}
\caption{Estimating the effect on $\rho_D^3$ and $\rho_{LS}^3$ of including
$1/m_b^{4,5}$ in the forward-backward asymmetry using Eq.~\eqref{Eq:DeltaHQP}
as a function of $E_\text{cut}$. }\label{Fig:RHODRHOLSdependence}
\end{figure}

The results for $\mathcal{M}$ being the forward-backward asymmetry $A_{FB}$ for the two parameters in $1/m_b^2$ are plotted in
Fig.~\ref{Fig:MUPIMUGdependence}, while the effect on the two parameters in
$1/m_b^3$ are plotted in Fig.~\ref{Fig:RHODRHOLSdependence} as a function of
the minimum charged lepton energy cut.

We find, that for increasing cut the effect for each of the non-perturbative
parameters is getting very small. The situation for a small cut, however, is
different. We would deem the effects on $\mu_\pi^2, \rho_D^3$ and especially on
$\rho_{LS}^3$ as significant, and too large. This indicates, that we are not in
particular sensitive to those heavy quark parameters. The situation for $\mu_G^2$ is
a bit different. Here the shift is in a reasonable order of magnitude and the
effect with a large energy cut is the largest, while the dependence on the charged lepton energy cut is the smallest. This more stable situation
confirms our previous finding, that we are most sensitive to $\mu_G^2$.
It also reflects the fact, that higher-order terms are getting less important
for an increasing cut on the charged lepton energy, which is contrary to naive
expectations.

\section{Discussion}\label{sec:discussion}

We have investigated inclusive semi-leptonic $B \rightarrow X_c \ell
\bar{\nu_\ell}$ decays in the context of heavy quark expansion, especially
with focus on a new observable. Our proposal is to utilise the forward-backward
asymmetry of the charged lepton as an additional constraint in measurements to
obtain information about the heavy quark parameters. 

First we have derived the triple differential decay rate in
Eq.~\eqref{Eq:TripleRate} including phase-space effects due to a minimum energy
cut on the charged lepton energy, which is required experimentally. After
revisiting the specific application of the heavy quark expansion in this case,
we construct the observable.

In section~\ref{sec:afbanalytic}, we have analysed the properties of the
forward-backward asymmetry without a cut at lower orders and compared to the
already existing observables, i.e. moments of the hadronic invariant mass and
charged lepton energy. It turns out, that we are specifically sensitive to the
$1/m_b^2$ corrections, and the linear combination of $\mu_\pi^2$ and $\mu_G^2$
is very similar to the total rate, from which $|V_{cb}|$ is finally extracted.
Hence we expect a large sensitivity to the chromo-magnetic moment $\mu_G^2$,
which can currently not be constrained very well from experimental analysis in
this decay mode.

Following up, we have investigated the full corrections in
section~\ref{sec:afbnumerical}, numerically. First we had a closer look onto
the differential spectrum in $z = \cos \theta$. It turns out, that in this
spectrum a cut on the charged lepton energy induces a discontinuity, that is
related to the hadronic system. This discontinuity is smoothed out by a
finite mass distribution of the hadronic system in reality. As this fact
reflects the dependence on the hadronic system in this variable, it is not
advisable to use this spectrum as an observable. However, integrated rates do
not suffer from this issue.

We therefore have analysed the forward-backward asymmetry $A_{FB}$. As said,
it is sensitive to the $1/m_b^2$ corrections, and especially it seems, that
$\mu_G^2$ may be constraint reasonably well from this for the first time only
due to
this decay analysis. It is a good candidate for an additional observable, that
will help to validate the heavy quark expansion, and at the same time increase
the precision on the extraction of $|V_{cb}|$.

In the numerical analysis it turns out, that the $1/m_b^4$ corrections seem to
be particularly large, while the corrections including $1/m_b^5$ are stable and
approach the results known from $1/m_b^3$ more closely. That might be related
to the occurrence of intrinsic charm operators, that mix the different orders
in power-counting starting at $1/m_b^4$ \cite{Bigi:2009ym}.

In future, one can study if New Physics operators, e.g. right-handed
currents, have a larger impact on this observable and hence may be constraint
in a better way. 

Furthermore one could in principle study a combined charged electron energy and/or hadronic invariant mass moment and $A_{FB}$
analysis, provided that this is experimentally feasible. For a generic observable, which combines $A_{FB}$ and a moment in the kinematic variable $\mathcal{M}$ we define
\begin{equation}
    \langle \mathcal{M} \rangle_{A_{FB}} =  \frac{ \int_{-1}^0 \text{d} z   \,\int\,\text{d}\,\mathcal{M}\,\frac{\text{d}^2 \Gamma}{\text{d}z\,\text{d}\mathcal{M}} \mathcal{M} - \int_{0}^1 \text{d} z  \,\int\,\text{d}\,\mathcal{M}\, \frac{\text{d}^2 \Gamma}{\text{d}z\,\text{d}\mathcal{M}} \mathcal{M} }{\int_{-1}^1 \text{d} z  \,\int\,\text{d}\,\mathcal{M}\,\frac{\text{d}^2 \Gamma}{\text{d}z \,\text{d}\mathcal{M}}}\,.
\end{equation}
This combination induces of course correlations with the other observables, but it may be sensitive to higher dimensional HQE parameters, which are not accessible right now. Evaluating this observable from Eq.~\eqref{Eq:FullContract}, this corresponds to moments of the hadronic structure function $W_3$, which again are not taken into account in current analysis.
For the predictions of
the charged lepton energy moment $\langle E_\ell \rangle_{A_{FB}}$ 
combined
with the forward-backward asymmetry, one needs to weight the integral over the
triple differential rate  with a factor of 
\begin{equation}
  \mathcal{M} = E_\ell = \frac12 \big(v{\cdot}q - z\sqrt{v{\cdot}q^2-q^2} \big)=\frac12
\big(m_b-v{\cdot}Q - z\sqrt{v{\cdot}Q^2-Q^2} \big)\,,
\end{equation}	       
see Eq.~\eqref{Eq:PhaseSpace}, while for the hadronic invariant mass $\langle M_X^2 \rangle_{A_{FB}}$ moment,  we need
the proper linear combination of moments in $v{\cdot}q$ and $q^2$
\begin{equation}
  \mathcal{M} = M_X^2 = M_B^2 - 2 M_B v{\cdot} q + q^2 = (M_B-m_b)^2 + 2(M_B-m_b) v{\cdot}Q +
Q^2\,.
\end{equation}
As higher orders in $M_X^2$ are in particular sensitive to $1/m_b$
correction terms in the expansion, see Eq.~\eqref{Eq:OT}, we expect that the
first moment in $M_X^2$ for the $A_{FB}$ potentially has an enlarged
sensitivity to $\rho_{LS}^3$, see the discussion around
Fig.~\ref{Fig:MUPIMUGdependence} and \ref{Fig:RHODRHOLSdependence}.

The achievable experimental uncertainties depend very much on the precision of the neutrino momentum reconstruction. For inclusive analysis, where the hadronic final state is not specified, this is obviously worse than for exclusive final states. However as we are interested in the normalised forward-backward asymmetry, only, and not in the angular spectrum,  hopefully most of the systematic uncertainties drop out. A remaining issue will probably be the migration of reconstructed events around the separations, i.e. $z=0$ and $z=z_\text{cut}$. A careful experimental analysis is required in order to assess an achievable precision for these observables.

In summary we have proposed a new observable for the analysis of semi-leptonic
$B\rightarrow X_c \ell \bar {\nu_\ell}$ decays. We have shown, that this
observable is indeed useful and have calculated the non-perturbative
corrections up to ${\cal O}(1/m_b^5)$.

\acknowledgments
S.T.  is supported by the ERC Advanced Grant EFT4LHC of the European Research
Council and the Cluster of Excellence Precision Physics, Fundamental
Interactions and Structure of Matter (PRISMA-EXC 1098). S.T. is grateful to Florian Bernlochner for useful discussions and comments about this manuscript. The author would like to
express special thanks to the Mainz Institute for Theoretical Physics (MITP)
for its hospitality and support during the workshop ''Challenges in
Semileptonic $B$ decays``, where part of this work has been developed and
discussed.

\appendix

\section{Full Analytic Results to Order ${\cal O}(1/m_b^3)$} \label{sec:appendixA}
The total rate is given by
\begin{align}
  \Gamma = \frac{G_F^2 | V_{cb}|^2}{192 \pi^3 m_b^5} \Big[&  1 -8 \rho -12 \rho ^2 \log\rho +8 \rho ^3 -\rho ^4  \nonumber \\
		      &-\frac{\mu_\pi^2}{2 m_b^2} \Big(  1 -8 \rho -12 \rho ^2 \log\rho +8 \rho ^3 -\rho ^4 \Big ) \nonumber \\
		      &+\frac{\mu_G^2}{2 m_b^2} \Big(  -3 +8 \rho  -12 \rho ^2 \log \rho -24 \rho ^2 +24 \rho ^3 -5 \rho ^4  \Big) \nonumber \\
		       &+\frac{\rho_D^3}{6 m_b^3} \Big( 77+48 \log (\rho)-88 \rho +36 \rho ^2 \log  \rho  +24 \rho ^2-8 \rho ^3 -5 \rho ^4 \Big) \nonumber \\
		       &-\frac{\rho_{LS}^3}{2 m_b^3} \Big(  -3 +8 \rho  -12 \rho ^2 \log \rho -24 \rho ^2 +24 \rho ^3 -5 \rho ^4  \Big) + {\cal O} \Big(\frac{1}{m_b^4}\Big )\Big]\,.\label{Eq:TRmb3}
\end{align}
The forward-backward asymmetry is given by
\begin{align}
  \Gamma  { \cdot} A_{FB} = &\frac{1}{4}\Big (1-20 \rho-90 \rho ^2 -20 \rho ^3+\rho ^4 +64 \rho ^{3/2} + 64 \rho ^{5/2}  \Big ) \nonumber \\
	&-\frac{\mu_\pi^2}{24 m_b^2} \Big( 35 -192 \sqrt{\rho } +420 \rho  + 210 \rho ^2 - 28 \rho ^3 + 3 \rho ^4 - 448 \rho ^{3/2} \Big ) \nonumber \\
	&+\frac{\mu_G^2}{24 m_b^2} \Big(   -65 +192 \sqrt{\rho } -60 \rho +330 \rho ^2-92 \rho ^3 +15 \rho ^4-320 \rho ^{3/2}  \Big ) \nonumber \\
	&+\frac{\rho_D^3}{24 m_b^3} \Big(  -35 +128 \sqrt{\rho } -140 \rho  +70 \rho ^2 -28 \rho ^3 +5 \rho ^4   \Big ) \nonumber \\
	&-\frac{\rho_{LS}^3}{8 m_b^3} \Big(  5 - 20 \rho + 30 \rho ^2 - 20 \rho ^3  + 5 \rho ^4    \Big ) + {\cal O} \Big(\frac{1}{m_b^4}\Big )\,.\label{Eq:AFBmb3}
\end{align}
The first charged lepton energy moment is given by
\begin{align}
 \Gamma {  \cdot}  \langle E_\ell \rangle &= \frac{m_b}{20}  \left(3 \rho ^5-15 \rho ^4+200 \rho ^3-60 \rho ^3 \log  \rho -120 \rho ^2-180 \rho ^2 \log  \rho -75 \rho +7 \right) \nonumber \\
	  &\phantom{= }\,+\frac{\mu _G^2}{6 m_b} \left( 3 \rho ^5-14 \rho ^4+24 \rho ^3-12 \rho ^2+5 \rho -12 \rho  \log  \rho -6\right) \nonumber \\
	  &\phantom{= }\,+\frac{\rho _D^3}{45 m_b^2} \left(18 \rho ^5-35 \rho ^4-120 \rho ^3+540 \rho ^2-730 \rho -120 \rho  \log  \rho +180 \log  \rho +327 \right)\nonumber \\ 
	  &\phantom{= }\,-\frac{3 \rho _{\text{LS}}^3}{5 m_b^2} \left( -\rho ^5+5 \rho ^4-10 \rho ^3+10 \rho ^2-5 \rho +1\right)  + {\cal O} \Big(\frac{1}{m_b^4}\Big )\,.\label{Eq:ELmb3}
\end{align}
The first hadronic invariant mass moment is given by
\begin{align}
\Gamma   {\cdot} \langle M_X^2 \rangle &= m_b M_B \left(-\frac{7 \rho ^5}{10}+\frac{9 \rho ^4}{2}-32 \rho ^3+6 \rho ^3 \log  \rho +16 \rho ^2+30 \rho ^2 \log  \rho +\frac{27 \rho
   }{2}-\frac{13}{10}\right)   \nonumber \\
   &\phantom{= }\,+m_b^2 \left(-\frac{3 \rho ^5}{10}+\frac{9 \rho ^4}{2}+24 \rho ^3-18 \rho ^3 \log  \rho -24 \rho ^2-18 \rho ^2 \log  \rho -\frac{9 \rho
   }{2}+\frac{3}{10}\right) \nonumber \\
   &\phantom{= }\,+M_B^2 \left(-\rho ^4+8 \rho ^3-12 \rho ^2 \log  \rho -8 \rho +1\right)   \nonumber \\ 
   &\phantom{= }\,+ \frac{\mu_\pi^2}{m_b^2} \Big[ m_b^2 \left(\frac{3 \rho ^5}{20}-\frac{9 \rho ^4}{4}-12 \rho ^3+9 \rho ^3 \log  \rho +12 \rho ^2+9 \rho ^2 \log  \rho +\frac{9 \rho }{4}-\frac{3}{20}\right)\Big]    \nonumber \\ 
   &\phantom{= }\,  +M_B^2 \left(\frac{\rho ^4}{2}-4 \rho ^3+6 \rho ^2 \log  \rho +4 \rho -\frac{1}{2}\right)    \nonumber \\ 
   &\phantom{= }\, + \frac{\mu_G^2}{m_b^2}  \Big[ m_b M_B \left(3-\frac{7 \rho ^5}{3}+\frac{35 \rho ^4}{3}-24 \rho ^3+\frac{52 \rho ^2}{3}+4 \rho ^2 \log  \rho -\frac{17 \rho }{3}+4 \rho  \log  \rho \right)  \nonumber \\ 
   &\phantom{= }\, +m_b^2
	\left(-\frac{3 \rho ^5}{4}+\frac{79 \rho ^4}{12}-4 \rho ^3-9 \rho ^3 \log  \rho +3 \rho ^2 \log  \rho -\frac{7 \rho }{12}-4 \rho  \log (\rho
	  )-\frac{5}{4}\right)  \nonumber \\ 
   &\phantom{= }\, +M_B^2 \left(-\frac{5 \rho ^4}{2}+12 \rho ^3-12 \rho ^2-6 \rho ^2 \log  \rho +4 \rho -\frac{3}{2}\right) \Big]      \nonumber \\ 
   &\phantom{= }\,  + \frac{\rho_D^3}{m_b^3}  \Big[ M_B^2 \left(-\frac{5 \rho ^4}{6}-\frac{4 \rho ^3}{3}+4 \rho ^2+6 \rho ^2 \log  \rho -\frac{44
	  \rho }{3}+8 \log  \rho +\frac{77}{6}\right)  \nonumber \\ 
   &\phantom{= }\,+ m_b M_B\! \left(\!-\frac{28 \rho ^5}{15}+\frac{44 \rho ^4}{9}+\frac{16 \rho ^3}{3}-\frac{112 \rho ^2}{3}+\frac{532 \rho }{9}+\frac{16}{3} \rho  \log\rho -16 \log \rho
      -\frac{452}{15}\right)    \nonumber \\ 
	  &\phantom{= }\,+m_b^2 \big(-\frac{\rho ^5}{4}-\frac{41 \rho ^4}{36}-4 \rho ^3+9 \rho ^3 \log  \rho +36 \rho ^2-23 \rho ^2 \log  \rho   \nonumber \\ 
   &\phantom{= }\,  -\frac{1831 \rho
	}{36}+\frac{8}{3} \rho  \log  \rho +8 \log  \rho +\frac{81}{4}\big)      \Big]  \nonumber \\ 	
   &\phantom{= }\, + \frac{\rho_{LS}^3}{m_b^3}  \Big[ \left(\frac{14 \rho ^5}{5}-14 \rho ^4+28 \rho ^3-28 \rho ^2+14 \rho -\frac{14}{5}\right) m_b M_B   \nonumber \\ 
   &\phantom{= }\, +m_b^2 \left(\frac{3 \rho ^5}{4}-\frac{79 \rho ^4}{12}+4 \rho ^3+9 \rho
   ^3 \log  \rho -3 \rho ^2 \log  \rho +\frac{7 \rho }{12}+4 \rho  \log  \rho +\frac{5}{4}\right)     \nonumber \\ 
   &\phantom{= }\, +M_B^2 \left(\frac{5 \rho ^4}{2}-12 \rho ^3+12 \rho ^2+6 \rho ^2 \log
    \rho -4 \rho +\frac{3}{2}\right) \Big] + {\cal O} \Big(\frac{1}{m_b^4}\Big )  \,.\label{Eq:MXmb3}
\end{align}

\end{document}